\documentclass[a4paper,12pt]{article}
\usepackage{amsmath,amssymb}
\usepackage{ifpdf}
\ifpdf
  \usepackage[pdftex]{graphicx}
\else
 \usepackage{ascmac}
\fi
\makeatletter
    
    \@addtoreset{equation}{section}
  \makeatother

\usepackage{here}

\newcommand{\cv}{{\cal V}}

\usepackage{color}

\usepackage{mathrsfs}

\usepackage{cite}

\makeatletter
\long\def\@makecaption#1#2{
  \vskip\abovecaptionskip
  \iftdir\sbox\@tempboxa{#1\hskip1zw#2}%
    \else\sbox\@tempboxa{#1~ #2}% 
  \fi
  \ifdim \wd\@tempboxa >\hsize% 
    \iftdir #1\hskip1zw#2\relax\par
      \else #1~ #2\relax\par\fi%
  \else
    \global \@minipagefalse
    \hbox to\hsize{\hfil\box\@tempboxa\hfil}% 
  \fi
  \vskip\belowcaptionskip}
\makeatother%% 
\newcommand{\eref}[1]{(\ref{#1})}

\newsavebox{\boxa}

\begin{document}

%%%%%%%%%%%%%%%%%%%%%%%%%%%%%%%%%%%%%%%%%%%%
\thispagestyle{empty}
\begin{flushright}
YITP-14-64, OU-HET-822
\end{flushright}
\vskip2cm
\begin{center}
{\Large {\bf %Generalization of Gradient Flow Equation and Its Application to Super Yang-Mills Theory
Generalized Gradient Flow Equation and Its Application to Super Yang-Mills Theory
% Gradient Flow Equation of Super Yang-Mills Theory
% Generalization of Gradient Flow Equation and Its Supersymmetric Extension
}}
\vskip1cm %\today\\
\vskip1cm
{\large
{Kengo Kikuchi$^1$\footnote{kengo@yukawa.kyoto-u.ac.jp}, Tetsuya Onogi$^2$\footnote{onogi@phys.sci.osaka-u.ac.jp} 
}

}
\vskip.7cm
{\it $^1$Yukawa Institute for Theoretical Physics, Kyoto University, Kyoto 606-8502, Japan  
\\
$^2$Department of Physics, Osaka University, Toyonaka 560-0043, Japan}\\

%%%%%%%%%%%%%%%%%%%%%%%%%%%%%%%%%%%%%%%%%%%%
\vskip1cm
\begin{abstract}
We generalize the gradient flow equation for field theories with nonlinearly realized symmetry.
%We give a gradient flow equation based on superfield formalism. 
Applying the formalism to super Yang-Mills theory,  we construct a supersymmetric extension of the gradient flow equation.
It can be shown that 
%both the supersymmetry and 
the super gauge symmetry is preserved in the gradient flow. 
Furthermore, choosing an appropriate modification term to damp the gauge degrees of freedom, 
we 
%also 
obtain a gradient flow equation which is closed within the Wess-Zumino gauge.
%Our result also provide the gradient flow equation of the matter field very naturally.
%Our result could provide a hint for the gradient flow equation of the matter field. 
\vspace{2.3cm}\vskip2cm

\end{abstract}

%%%%%%%%%%%%%%%%%%%%%%%%%%%%%%%%%%%%%%%%%%

%%%%%%%%%%%%%%%%%%%%%%%%%%%%%%%%%%%%%%%%%%%%

\end{center}

%%%%%%%%%%%%%%%%%%%%%%%%%%%%%%%%%%%%%%%%%%

\section{Introduction}\label{introduction}
In recent years, the gradient flow equation has been the focus of much attention. 
The equation was proposed by Martin Luscher \cite{Luscher:2010iy} for Yang-Mills theory as a certain type of 
diffusion equation to give a one parameter deformation of the gauge field evolving in the flow time 
starting from the bare gauge field as the initial condition. It was found \cite{Luscher:2011bx}
that the expectation value of any gauge 
invariant local operators of the new gauge field, which is the solution of the gradient flow equation, is finite 
without additional renormalization.

Various applications of the physical observable are studied recently. Ref.~\cite{Luscher:2013vga} give a review 
of the recent applications.
For example, the gradient flow of a matter field $\chi$ is proposed as follows~\cite{Luscher:2013cpa}:
\begin{eqnarray}
\dot{\chi}=\Delta\chi,~~~\chi |_{t=0}=\psi, \label{fundamentalluscher}\\
\Delta=\ooalign{\hfil/\hfil\crcr $D$}^2~ \mathrm{or~simply}~\Delta=D_{\mu}D_{\mu},\label{fundamentalluscher2}
\end{eqnarray}
where $D_\mu=\partial_{\mu}+B_{\mu}$. Using this equation, the expectation value of the chiral densities is calculated \cite{Luscher:2013cpa}. 
The relation between the small flow time behavior of certain gauge invariant local products and the correctly-normalized conserved energy-momentum tensor in the Yang-Mills theory is given~\cite{Suzuki:2013gza}. More appropriate probes for the translation Ward identities is defined \cite{DelDebbio:2013zaa}. 
The methods is also applied in the lattice theory \cite{Luscher:2010iy, Fodor:2012td, Fodor:2012qh, Fritzsch:2013je, Fritzsch:2013hda, Ramos:2013gda, Asakawa:2013laa, Luscher:2014kea, Rantaharju:2013bva, Makino:2014wca, Makino:2014taa, Fritzsch:2013yxa, Bar:2013ora, Brida:2013mva, Monahan:2013lwa, Shindler:2013bia}, a new scheme of the step scaling, the improved action, and so on. 

In this way, the gradient flow equation has spurred a great deal of research. 
%One interesting question is to what other physical system this method can be applied. 
In view of this nice property, it is natural to consider possible extensions of this method for other theories.
%whether the method of gradient flow,  when applied to other field theories, can retain the nice property on the renormalization. 
%The equation is very attractive, therefore it is worth extending the equation to other theory, for example, to the QCD with matter field.
%This has lead to the application of the method of gradient flow QCD with matter field by Luscher as Eq.~\eref{fundamentalluscher} . 
%One expects that analogy with the Yang-Mills theory might help to construct the gradient flow equation.
%However, this is not the case in general when there are fields with different mass dimensions such as 
%fermion fields.
%
%However the quark part of this equation is no longer defined from the gradient of the action. We would like to derive the gradient flow equation of matter fields more systematically, that is, we would like to obtain the gradient flow equation of the matter field from the gradient of an action.
One interesting system 
%along this line 
is the super Yang-Mills theory.  This theory shares similar property as the Quantum Chromodynamics (QCD) since it has a  matter field called gaugino, though it is in the adjoint representation. On the other hand,  
the restriction from the supersymmetry (SUSY) can give a natural extension of the gradient flow in Yang-Mills theory.
In this sense, super Yang-Mills theory could be an interesting laboratory for studying the extension of the gradient flow equation. \\
Once it is constructed, there can be interesting applications of the supersymmetric Yang-Mills lattice theory. 
For example,  imposing the supersymmetric Ward-Takahashi identity for composite fields with finite flow time, 
one might be able to  determine the renormalization factor for the super current as well as various improvement terms in the action and operators, in close analogy with Luscher's  work on the chiral symmetry  in QCD with Wilson fermions.
This method may also be useful for testing the validity of various proposals of supersymmetric lattice models.

%\textcolor{blue}{
%Of course, since the super Yang-Mills theory is an attractive for its own sake, it would be useful to study the gradient flow 
%equation for this system. Also, it may give us a hint to understand the theoretical basis of the gradient flow equation including 
%matter fields. }

In this paper, we construct the gradient flow equation for super Yang-Mills theory. Since the super gauge symmetry is nonlinearly realized,  
we first construct the generalization of the gradient flow equation for quantum field theories with nonlinearly realized symmetry.
Applying the generalized equation to super Yang-Mills theory, we construct a natural extension of the gradient flow using superfield formalism. We find that  with a special choice of the modification term in the gradient flow equation, we obtain a closed equation 
within  the Wess-Zumino (WZ) gauge.  %Our result can give a hint for the derivation of the gradient flow equation for the matter field.
%Also we study the gradient flow of the matter field. Luscher propose the gradient flow of the matter field in Ref.~\cite{Luscher:2013cpa}, but there is room for further research to derive the equation for matter field. 
%Since the super Yang-Mills theory contains gaugino as a `matter' field, 

This paper is organized as follows. In Sec.~\ref{g1}, we give a brief review of the method of gradient flow in Yang-Mills theory. In Sec.~\ref{s0} we propose the generalized gradient flow equation of the quantum field theory with nonlinearly realized symmetry.  In Sec.~\ref{s1} we apply the generalized equation to super Yang-Mills theory. To obtain the compact expression of the equation of super Yang-Mills theory, we look for a special choice of the modification term to give a closed equation within the WZ gauge. In Sec.~\ref{SUSY Non-Abelian}, we give the gradient flow equation for super Yang-Mills theory concretely with component fields 
including the gaugino field. Sec.~\ref{summary3} is devoted to summary and discussions. %We give our notation in Appendix \ref{appz}.
%We give our notation in Appendix \ref{appz}. 

%In Sec.~\ref{SUSY Abelian}, we extend the gradient flow equation to pure abelian SUSY theory as an example. 

%%%%%%%%%%%%%%%%%%%%%%%%%%%%%%%%%%%%%%%%%%%%%%%%%%%%%%%%%%%%%%%%%%%%%%%%%%

\section{Gradient Flow Equation for Yang-Mills Theory}\label{g1}
%As mentioned in the introduction, the analogy with the Yang-Mills theory can help to construct the gradient flow equation of  the super Yang-Mills theory.  For this purpose, let us review the Yang-Mills gradient flow equation first.

% In the method of the gradient flow equation, the expectation value in terms of new gauge field is finite without additional renormalization. Here the new gauge field  is constructed by the solution of a certain type of diffusion equation called a gradient flow equation, whose initial value is the bare gauge field.  Luscher showed this at one loop order as example \cite{Luscher:2010iy}, and later all order proof was given by Luscher and Weisz \cite{Luscher:2011bx}. %We call the statement the Luscher-Weisz theorem. 

%\subsection{Definition of Gradient Flow Equation}
The gauge field $B_\mu$ is defined by the gradient flow equation
\begin{eqnarray}
&&\dot{B}_\mu=D_\nu G_{\nu\mu}+\alpha_0 D_\mu\partial_\nu B_\nu,\label{1}\\
&&B_\mu |_{t=0}=A_\mu.\label{2121}
\end{eqnarray}
where the dot means a differential in terms of the flow time $t$, $A_\mu$ describe a fundamental bare field of $SU(N)$ gauge theory, $G_{\mu\nu}$ and $D_\mu$ are defined by
\begin{eqnarray}
G_{\mu\nu}&=&\partial_\mu B_\nu-\partial_\nu B_\mu+[B_\mu, B_\nu],\\
D_\mu&=&\partial_\mu+[B_\mu, \cdot ]
\end{eqnarray}
respectively. The reason why we call the equation the gradient flow one is that the first term of R.H.S. of Eq.~\eref{1} is proportional to  the gradient of the action,
\begin{eqnarray}
S=\int d^4 x \mathrm{Tr}[G_{\mu\nu}(x) G_{\mu\nu}(x)].\label{yangmillsaction}
\end{eqnarray}
The second term of the R.H.S. of Eq.~\eref{1} is a modification term to damp the gauge degrees 
of freedom. In this paper, we call this term as the $\alpha_0$ term. It has to be proportional to the form of gauge transformation so that it does not affect the evolution of gauge invariant observables. 
Luscher claims that any expectation value which is described by the gauge field $B_\mu$, which is defined by Eq.~\eref{1} at positive flow time has a well-defined continuum limit without additional renormalization. Calculating the expectation value of the energy density at one loop order using this method, he showed that it is the case \cite{Luscher:2010iy}. Soon after that, Luscher and Weisz proved this claim to all order in perturbation theory \cite{Luscher:2011bx}. Hereafter, we call the claim of the all order finiteness of the observables constructed from the gauge field $B_\mu$ with finite flow time as ``the Luscher-Weisz theorem''. Eq.~\eref{1} has a gauge symmetry at any flow time if the gauge parameter $\omega(t, x)$ satisfies the condition
\begin{eqnarray}
\partial_t \omega=\alpha_0 D_\mu \partial_\mu \omega.
\end{eqnarray}
The symmetry of the equation is a key to prove the Luscher-Weisz theorem.

%Luscher also proposed the equation of the matter field as in Eqs.~\eref{fundamentalluscher} and \eref{fundamentalluscher2}. The way how to derive the gradient flow equation of the matter field, however, is not clear and also there is some arbitrariness in the choice of the flow equation. It would be nice if we can find a way to derive the flow equation in a systematic fashion. For this purpose, we focus on the super Yang-Mills theory. This theory, as is well known, contains the gaugino as a matter field. If we derive to the gradient flow equation of the vector superfield in the super Yang-Mills theory by the analogy of the Yang-Mills gradient flow, it could give us a hint for the derivation of the gradient flow equation of matter fields.

When we extend the method of gradient flow to the super Yang-Mills theory, we encounter a problem.
Since the super gauge symmetry is nonlinearly realized, the naive gradient flow equation does not respect 
the super gauge symmetry. In order to solve this problem, we propose a generalization of the gradient flow equation 
with nonlinearly realized symmetry in the next section.

%Using the gradient flow equation, 
%Luscher \cite{Luscher:2009eq}

%has proven this at one loop order \cite{Luscher:2010iy}, and later all order proof (We call it the Luscher-Weisz theorem) was given by Luscher and Weisz \cite{Luscher:2011bx}.

%\vspace{0.5cm}

%\begin{itembox}[l]{Luscher-Weisz Theorem} 
%In the Yang-Mills theory, the expectation value in terms of the new gauge field, which is defined by the solution of the gradient flow equation, are finite without additional renormalization to all loop order, once the theory in terms of the fundamental gauge field is renormalized in the usual way.
%\end{itembox}

%\vspace{0.5cm}
%There are various applications owing to the theorem as we shall explain later in Sec.~\ref{g2}.  

%%%%%%%%%%%%%%%%%%%%%%%%%%%%%%%%%%%%%%%%%%%%%%%%%%%%%%%

\section{General Form of Gradient Flow Equation}\label{s0}
The gradient flow gives the steepest descent time evolution in the space based on the 
energy function of the system. For example, when the space is $\mathbf{R}^n$, the gradient flow equation is given as
\begin{eqnarray}
\frac{dq^i}{dt} = - \frac{\partial E(q)}{\partial q^i} & (i=1,\cdots, n),
\end{eqnarray}
where $q^i (i=1,\cdots,n)$ is the position in $\mathbf{R}^n$ and $E(q)$ is the energy function.
The time evolution of the energy function is then given as
\begin{eqnarray}
\frac{dE(q)}{dt} = \sum_{i=1}^n \frac{\partial E(q)}{\partial q^i} \frac{dq^i}{dt} = -\sum_{i=1}^n \left(\frac{\partial E(q)}{\partial q^i} \right)^2 \leq 0,
\end{eqnarray}
so that the energy decreases monotonically in time towards an extremum of the energy function.

The gradient flow can be naturally extended to field theory. For example, in the field theory in $4$ dimension with field $\phi(x)$ with action $S(\phi)$, taking the space 
$\{\phi(x)\}$ as the functional space and the action $S(\phi)$ as the energy function the gradient flow equation becomes
\begin{eqnarray}
\frac{\partial \phi_t(x)}{\partial t} = - \frac{\delta S(\phi_t)}{\delta \phi_t(x)}.
\label{eq:naive_GF}
\end{eqnarray}
Since the flow stops when the field reaches the extremum of the action, the gradient flow gives an interpolation of the initial field and a classical solution of the theory.
%\textcolor{red}{\it is this necessary? is there better explanation?}
% In the method of the gradient flow equation, the expectation value in terms of new gauge field is finite without additional renormalization. Here the new gauge field  is constructed by the solution of a certain type of diffusion equation called a gradient flow equation, whose initial value is the bare gauge field.  Luscher showed this at one loop order as example \cite{Luscher:2010iy}, and later all order proof was given by Luscher and Weisz \cite{Luscher:2011bx}. %We call the statement the Luscher-Weisz theorem. 
%\subsection{Definition of Gradient Flow Equation}

There is a question whether the symmetry of the theory is preserved under the time evolution 
with the gradient flow equation of the type in Eq.\eref{eq:naive_GF}. Fortunately, in Yang-Mills theory case,  the gauge symmetry is preserved. 
However, in general, the problem can arise when the symmetry of the system is nonlinearly realized.

It turns out that the super Yang-Mills theory 
has a super gauge symmetry which is nonlinearly realized with respect to the vector superfield. 
Since the BRS symmetry is the key for Luscher-Weisz theorem for Yang-Mills theory, one can expect 
that the super gauge symmetry in super Yang-Mills theory could also play a crucial role. 
In the following subsections, we consider the field theories with a nonlinearly realized symmetry 
and construct the generalization of the gradient flow equation which respects the symmetry.

\subsection{Field Theories with Nonlinearly Realized Symmetry}
Let us now generalize the gradient flow for field theories with a symmetry which is nonlinearly realized.
\begin{eqnarray}
\phi^a(x)  \rightarrow \phi^{\prime a}(x), &&  (a=1,\cdots, M)
\end{eqnarray}
where $a$ is the index for the internal degrees of freedom and $M$ is the total number of components. 
 $\phi^\prime$ is a nonlinear function of $\phi$. 
 Under this transformation, the action is invariant
\begin{eqnarray}
S(\phi^\prime)=S(\phi).
\label{eq:S_symmetry}
\end{eqnarray}
One can find that the naive gradient flow equation given in Eq.(\ref{eq:naive_GF}) does not keep the symmetry. 
This is because the gradient flow equation based on Eq.(\ref{eq:naive_GF}) for the field after the transformation $\phi^\prime$ reads
\begin{eqnarray}
\frac{\partial \phi^{a \prime}_t(x)}{\partial t} = - \frac{\delta S(\phi^\prime_t)}{\delta \phi^{a \prime}_t(x)}.
\end{eqnarray}
Using Eq.(\ref{eq:S_symmetry}) and the chain rule for derivative, the above equation becomes
\begin{eqnarray}
\sum_{b=1}^{M} \frac{\partial \phi^{a \prime}_t(x)}{\partial \phi^b_t(x)} \frac{\partial \phi^b_t(x)}{\partial t} 
= - \sum_{b=1}^M \frac{\partial \phi^b_t(x)}{\partial  \phi_t^{a \prime}(x)}\frac{\delta S(\phi_t)}{\delta \phi^b_t(x)}.
\end{eqnarray}
Multiplying $\displaystyle{\frac{\partial\phi^c_t(x)}{\partial \phi_t^{a \prime}(x)}}$ and sum over $a$, 
the naive gradient flow equation for $\phi_t^\prime$ would reduce to
\begin{eqnarray}
 \frac{\partial \phi^c_t(x)}{\partial t} 
= -\sum_{a=1}^M 
\frac{\partial \phi^c_t(x)}{\partial \phi_t^{a \prime}(x)}
\frac{\partial \phi^b_t(x)}{\partial \phi_t^{a \prime}(x)}
\frac{\delta S(\phi_t)}{\delta \phi^b_t(x)}.
\end{eqnarray}
Thus the naive gradient flow equations before and after the symmetry transformation are different.

 \subsection{Our Proposal for Generalized Gradient Flow Equation}
%\textcolor{red}{\it Whenever the word "configuration" appear, change it to "functional".}\\
How can we define a gradient flow equation which respects the symmetry?
As we have seen in the previous subsection, the problem in the naive gradient flow equation 
is that the L.H.S. and the R.H.S. transform differently under the symmetry transformation.
A natural solution could be to introduce a ``metric'' to compensate the mismatch of the transformation 
property. The metric in functional space for the field theory in $D$ dimension can be defined through the 
norm of the variation of fields $\delta \phi(x)$ which is invariant under the symmetry.
\begin{eqnarray}
||\delta \phi||^2=\int d^D x g_{ab} (\phi(x))\delta \phi^a (x) \delta \phi^b (x),~~~~~a=1, 2,\cdots, M
\label{eq:norm_phi}
\end{eqnarray}
where $M$ is the number of components of the field and $g_{ab}(\phi(x))$ is the metric in the functional space.
The metric should be chosen in such a way that the norm is invariant under the symmetry transformation as
\begin{eqnarray}
||\delta \phi^\prime||^2 =||\delta \phi ||^2,
\end{eqnarray}
which leads to the following properties for the metric in the functional space.
\begin{eqnarray}
g_{ab}(\phi^\prime(x))&=&
\displaystyle{\frac{\partial \phi^c(x)}{\partial \phi^{\prime a}(x)}
\frac{\partial \phi^d(x)}{\partial \phi^{\prime b}(x)}}
 g_{cd}(\phi(x)), 
\label{eq:isometry1}
 \\
 g^{ab}(\phi^\prime(x))&=&
\displaystyle{\frac{\partial \phi^{\prime a}(x)}{\partial \phi^c(x)}
\frac{\partial \phi^{\prime b}(x)}{\partial \phi^d(x)}}
 g^{cd}(\phi(x)).
\label{eq:isometry2}
 \end{eqnarray}
 Whether one can find an appropriate metric or not for a given field theory is quite nontrivial, 
 but there are quite a few examples  in which one can find the metric explicitly 
 such as $O(N)$ nonlinear sigma model or $SU(N)$ lattice gauge theory.
 In any case,  Eqs.~(\ref{eq:isometry1}) and (\ref{eq:isometry2}) mean that symmetry transformation is the isometry for the metric defined through the invariant norm. 
 
%\textcolor{red}{\it remove item box everywhere!!}
The condition for the isometry in Eq.(\ref{eq:isometry2}) gives exactly the right quantity to compensate the mismatch of the transformation 
property in the naive gradient flow equation. Thus we find that when we require the invariance of the gradient flow under the symmetry,  
the gradient flow should be modified as 
%\begin{itembox}[l]{General form of Gradient Flow Equation}
\begin{eqnarray}
\frac{\partial \phi_t^a(x)}{\partial t}=-g^{ab}(\phi_t(x))\frac{\delta S(\phi_t)}{\delta \phi_t^b(x)} \  .
\label{eq:GGFloweq}
\end{eqnarray}
%\end{itembox}\vspace{0.5cm}
%Since both the L.H.S. and the R.H.S. of the equation are contra-variant vectors, the equation is manifestly covariant under 
%symmetry transformation.
% \textcolor{red}{\it we do not understand what we mean by isometry very well. change it to more 
%usual expression such as "symmetry is manifestly preserved.."}
%In order to understand that our proposal does give a gradient flow which is consistent with the symmetry, 
%let us consider the symmetry transformation of the system. 
In what follows, we call the above equation as the generalized gradient flow equation. It is clear  the time evolutions of $\phi$ and $\phi^\prime$ fields with our generalized 
gradient flow equation are mutually consistent under symmetry transformation
The evolution equation \eref{eq:GGFloweq} was also discussed in the context of Fokker-Plank equation.
\cite{graham1, graham2, Namiki:1984xe, Rumpf:1985eh, ZinnJustin:1986eq, Halpern:1987ub, Nakazawa:1988ss, Nakazawa:1989jf}.
%\subsection{Examples}

In Appendices \ref{NLS} and \ref{LGT}, we apply the generalized gradient flow equation to the $O(N)$ nonlinear sigma model and the $SU(N)$ lattice gauge theory and verify its validity. We find that the generalized gradient flow equation gives a time evolution 
which respects the nonlinearly realized symmetry of the system. 
In the next section, we construct the gradient flow equation of the super Yang-Mills theory based on 
the generalized gradient flow equation which respects the super gauge symmetry.

\section{Supersymmetric Gradient Flow Equation}\label{s1}\label{proposal}

\subsection{Derivation of Gradient Flow Equation of Super Yang-Mills Theory}
%We success the expansion of the gradient flow equation to super Yang-Mills theory. 
Before studying the super Yang-Mills theory, let us review the steps for constructing the gradient flow equation 
in ordinary non-SUSY Yang-Mills theory. The local gauge transformation is given as 
\begin{eqnarray}
A_\mu(x) \rightarrow A_\mu(x) + D_\mu \omega(x),
\end{eqnarray}
where $D_\mu$ is the covariant derivative and $\omega(x)$ is the gauge transformation parameter.
One can see that the invariant norm of the vector field $\delta A_\mu(x)$ is given as 
\begin{eqnarray}
||\delta A_\mu(x)||^2 = \int d^4x \mathrm{Tr}\left[ \delta A_\mu(x) \delta A_\mu(x)\right], 
\end{eqnarray}
which means that the metric in the field space is 
\begin{eqnarray}
g^{ab}(A_\mu) = 2\delta^{ab} 
\end{eqnarray}
Therefore, 
how to derive the gradient flow in ordinary Yang-Mills theory can be summarized as follows:
\begin{enumerate}
\item Starting from the Yang-Mills action $S_{\mathrm{YM}}$, we make a variation over the $A^b_\mu(x)$ field, and multiply the metric $2\delta^{ab}$, where
\begin{eqnarray}
S_{\mathrm{YM}}&=&\int d^4 x \mathrm{Tr}[F_{\mu\nu}(x)F_{\mu\nu}(x)],\\
F_{\mu\nu}&=&\partial_\mu A_\nu -\partial_\nu A_\mu+[A_\mu, A_\nu].
\end{eqnarray}
\item We replace the $A_\mu(x)$ field with the new gauge field $B_\mu(t, x)$, and impose the initial condition $B_\mu(0, x)=A_\mu(x)$, and introduce the field strength $G_{\mu\nu}\equiv \partial_\mu B_\nu-\partial_\nu B_\mu+[B_\mu, B_\nu]$.
\item We add a new gauge fixing term to suppress the increase of the degree of new gauge freedom in the flow time direction.  It has to be proportional to the gauge transformation, because  physical quantities do not depend on the term.
\item We regard the sum of them as R.H.S.~of the gradient flow equation. 
\item We regard the derivative of $B^a_\mu(t, x)$ with respect to $t$ as L.H.S.~of the gradient flow equation.
\end{enumerate}
Thus, we obtain the gradient flow equation in Yang-Mills theory as Eqs.~\eref{1} and \eref{2121}.

We now apply the general gradient flow equation to super Yang-Mills theory. 
The super gauge transformation of the super Yang-Mills vector superfield $V$ is given as 
\begin{eqnarray}
e^V \rightarrow e^{-i\Lambda^\dagger}e^Ve^{i\Lambda},
\end{eqnarray}
where $\Lambda,\Lambda^\dagger$ are arbitrary chiral and anti-chiral superfields. The component of superfield $V$ is defined by $V=\{C,X,\bar{X},M,M^*,V_m,\Lambda, \bar{\Lambda},D\}$.
The invariant norm for $\delta V$ under the super gauge transformation is then given as 
\begin{eqnarray}
||\delta V||^2 \equiv -\int d^8z  \mathrm{Tr} \left[ e^{-V} \left(\delta e^V\right) e^{-V} \left(\delta e^V\right)\right].
\end{eqnarray}
This means that the space of vector superfields has a nontrivial metric in functional space.
To obtain the gradient flow equation of the super Yang-Mills theory, we replace the statement partly as follows:
\begin{itemize}
\item Yang-Mills action $S_{\mathrm{YM}}$ $\rightarrow$ Super Yang-Mills action $S_{\mathrm{SYM}}$,
where
\begin{eqnarray}
S_{\mathrm{SYM}}&=&-\int d^4 x \int d^2\theta \mathrm{Tr}[W^\alpha W_\alpha]+h.c.,\\
W_{\alpha}&=&-\bar{D}\bar{D}(e^{-V}D_{\alpha}e^V).
\end{eqnarray}
\item Gauge field $A_\mu(x)$  $\rightarrow$ Superfield $V(z)$. The argument $z$ stands for super coordinate $(x, \theta, \bar{\theta})$. 
\item New gauge field $B_\mu(t, x)$ $\rightarrow$ New superfield $\cv (t, z)$. The component of superfield $\cv$ is defined by ${\cal V}=\{c,\chi,\bar{\chi},m,m^*,v_m,\lambda, \bar{\lambda},d\}$. We impose the initial condition $\cv(0, z)=V(z)$.
\item Gauge transformation $\rightarrow$ Super gauge transformation.
\item Metric $g^{ab}(A_\mu)$ $\rightarrow$ $g^{ab}(V)$   
\end{itemize}
Thus we propose a general form of the supersymmetric extension of the gradient flow equation.
\begin{eqnarray}
\frac{\partial \cv^a}{\partial t}=-g^{ab}(\cv)\frac{\delta S_{\mathrm{SYM}}}{\delta \cv^b}+\alpha_0\delta \cv^a\label{general1}.
\end{eqnarray}
The $\delta \cv$ is the super gauge transformation of $\cv$, which is defined by the equation as
\begin{eqnarray}
\delta \cv=L_{\cv/2}\cdot [(\Phi-\Phi^{\dagger})+\mathrm{coth}(L_{\cv/2})\cdot(\Phi+\Phi^{\dagger})],
\end{eqnarray}
where $\Phi$ is a chiral superfield. 
%When we substitute the explicit form of $g^{ab}(v), \frac{\delta S_{\mathrm{SYM}}}{\delta v^b},  \delta v_a$, which are given in Appendix \ref{explicit}, we can obtain more explicit form as follows:
Substituting the explicit forms of $g^{ab}(\cv), \frac{\delta S_{\mathrm{SYM}}}{\delta \cv^b},  \delta \cv_a$, we obtain the gradient flow equation in the matrix form as 
\begin{eqnarray}
\frac{\partial \cv}{\partial t}%&=&\frac{L_v}{1-e^{-L_v}}F+\frac{L_v}{e^{L_v}-1}F^{\dagger}+\alpha_0\delta v\\
&=&\frac{L_\cv}{1-e^{-L_\cv}}(F+\alpha_0\Phi_\cv)+h.c.,\label{81}
\end{eqnarray}
where
\begin{eqnarray}
F=D^\alpha w_\alpha+\{e^{-\cv}D^\alpha e^{\cv}, w_\alpha\}.
\end{eqnarray}
and $\Phi_\cv$ is a chiral field, $\cv=\cv^a T^a$ and $T^a$ is a representation matrix. The field strength $w_\alpha$ is given by $w_\alpha \equiv -\bar{D}\bar{D}(e^{-\cv}D_\alpha e^{\cv})$. The $L_\cv $ is defined by
\begin{eqnarray}
L_\cv ~\cdot \equiv[\cv, ~\cdot~ ].\end{eqnarray}
The derivation of the above equation is given in Appendix \ref{explicit}. The covariant term of Eq.\eref{81} was also discussed in the stochastic quantization\cite{Nakazawa:2004ac, Nakazawa:2003tz}.
%Here we use the fact that the $\alpha_0$ term can be described by
%\begin{eqnarray}
%\alpha_0 \left(\frac{L_v}{1-e^{-L_v}}\Phi_v+h.c. \right)
%\end{eqnarray}
%using a chiral field $\Phi_v$. 
We can also rewrite Eq.~\eref{81} more simply using $e^\cv$ as a basic variable as
\begin{eqnarray}
\frac{\partial e^\cv}{\partial t}=e^\cv(F+\alpha_0 \Phi_\cv)+h.c..
\end{eqnarray}
This form is useful for studying the time dependence of the super gauge transformation as discussed in the next subsection.

\subsection{Symmetries of Gradient Flow Equation}

We comment on the supersymmetry and super gauge symmetry of the gradient flow equation. The equation consists of covariant derivative operators $D, \bar{D}$, and vector multiplet $\cv$. Because supersymmetric transformation operators $Q\xi, \bar{Q}\bar{\xi}$  commute with $D, \bar{D}$, the equation keeps SUSY manifestly if $\xi$ and $\bar{\xi}$ do not depend on the flow time.   

It is important to examine the condition that the gradient flow equation has super gauge symmetry at any flow time. Taking the infinitesimal super gauge transformation for both sides of the gradient flow equation, we obtain the 
condition for $\Lambda$, 
\begin{eqnarray}
i\frac{d \Lambda}{d t}&=&\alpha_0(\delta_{\Lambda} \Phi_\cv+i[\Lambda, \Phi_\cv])\label{cond1}\end{eqnarray}
The $\delta_\Lambda$ is infinitesimal super gauge transformation,
\begin{eqnarray}
\Phi_\cv&\rightarrow&\Phi_\cv+\delta_\Lambda\Phi_\cv
\end{eqnarray}
If $\Lambda$ satisfies the condition Eq.~\eref{cond1}, the gradient flow equation is invariant for the super gauge transformation at any flow time.

\section{Gradient Flow Equation of Super Yang-Mills Theory under Wess-Zumino Gauge}

In this section, we determine the form of the gradient flow equation of super Yang-Mills theory under the WZ gauge. Because Eq.~\eref{81} have infinite number of terms, it is very difficult to solve it. In order to obtain the flow equation with finite number of terms, we choose the WZ gauge. 

However, generally the time evolution from the flow equation can carry the system away from the WZ gauge. Therefore, the most important question is whether there exists the special chiral field $\Phi_\cv$ which give the super gauge transformation keeping the WZ gauge. As a result, we find that such a $\Phi_\cv$ exists. 

Here we discuss how to determine the form of the $\alpha_0$ term.
%The gauge transformation is defined by
%\begin{eqnarray}
%\delta V&=&L_{V/2}\cdot[(\Phi-\Phi^{\dagger})+\coth{(L_{V/2})}\cdot(\Phi+\Phi^{\dagger})],
%\end{eqnarray}
%where $\Phi$ and $\Phi^{\dagger}$  are a chiral superfield and an anti-chiral superfield respectively. Under the WZ gauge, the gauge transformation is expressed by a finite number of terms as
%\begin{eqnarray}
%\delta V&=&\Phi+\Phi^{\dagger}+\frac{1}{2}[V, \Phi-\Phi^{\dagger}]+\frac{1}{12}[V, [V,\Phi+\Phi^{\dagger}]].
%\end{eqnarray}
 We try to find out the special form of the $\alpha_0$ term so that the gradient flow equation is consistent within the WZ gauge. This means the $\alpha_0$ term has to satisfy the following requirements. 
\begin{itemize}
\item It is positive.
\item The mass dimension is two.
%\item It is chiral (or anti-chiral) field.
\item It is described by super gauge transformation $\delta \cv$.
\item The flow of the vector field keeps the WZ gauge at any flow time.
\end{itemize}
%It is possible that there are some $\Phi$ (or $\Phi^\dagger$) which satisfy the requirements. The most important thing , however, is  that there exist at least one such term. This is because the physical quantity does not depend on the gauge fixing. 
As a result,  we found out that there exists at least one example of the $\alpha_0$ term which satisfies these conditions.

\begin{eqnarray}
\alpha_0&=&1,\\
\delta \cv&=&\Phi_\cv+\Phi_\cv^{\dagger}+\frac{1}{2}[\cv, \Phi_\cv-\Phi_\cv^{\dagger}]+\frac{1}{12}[\cv, [\cv,\Phi_\cv+\Phi_\cv^{\dagger}]],\\
\end{eqnarray}
where\begin{eqnarray}
\Phi_\cv&=&\bar{D}^2(D^2\cv+[D^2 \cv, \cv]).\label{important}
\end{eqnarray}
It is possible that $\Phi_\cv$ which gives the super gauge transformation keeping the WZ gauge may not be unique. However, this example can be useful for further studies. %The most important things, however, is to exist at least one such the $\alpha_0$ term. 

%The derivation of this equation is described in Sec.~\ref{gaugefixing}. Using this new equation, we find out that, in this case, an extra term is required in addition to the gradient flow equation of the matter field which is proposed by Luscher.

%%%%%%%%%%%%%%%%%%%%%%%%%%%%%%%%%%%%%%%%%%%%%%%%%%%%%%
\section{Gradient Flow Equation of Super Yang-Mills Theory for Each Component}\label{SUSY Non-Abelian}\label{s3}

In this section, we applied our equation, which is obtained in Sec.~\ref{s1}, to super Yang-Mills theory concretely, and derive the gradient flow equation of each component under WZ gauge. It gives the gradient flow equation of the matter field. For the sake of understanding this section, we give the equation in the case of the pure Abelian supersymmetric theory in Appendix~\ref{SUSY Abelian} as an example.

\subsection{Expansion in Component Fields}
We rewrite $F$ for the convenience as
\begin{eqnarray}
F=D^\alpha w_\alpha+\{e^{-\cv}D^\alpha e^{\cv}, w_\alpha\}.\label{Aterm}
\end{eqnarray}
Useful formulae to expand Eq.~\eref{Aterm} in component fields are given in Appendix \ref{E}. The gauge covariant term is given as
\begin{eqnarray}
&&\hspace{1cm}\left( \frac{L_\cv}{1-e^{-L_\cv}}\cdot F \right) +\left( \frac{L_\cv}{e^{L_\cv}-1}\cdot F^{\dagger}\right)\nonumber\\
&=&F+F^{\dagger}+\frac{1}{2}[\cv, F-F^{\dagger}]+\frac{1}{12}[\cv, [\cv, F+F^{\dagger}]]+O(\cv^3),
\end{eqnarray}
where $F$ is represented in $(x, \theta, \bar{\theta})$ coordinates by
\begin{eqnarray}
F(x, \theta, \bar{\theta})&=&-8d+8\theta\sigma^m\mathscr{D}_m\bar{\lambda}-8\bar{\theta}\bar{\sigma}^m\mathscr{D}_m\lambda\nonumber\\
&&+4(\bar{\theta}\bar{\sigma}^m\theta)[v_m, d]+4(\theta\sigma^k\bar{\sigma}^m\sigma^l\bar{\theta})\mathscr{D}_l v_{mk}+8[\bar{\theta}\bar{\lambda}, \theta\lambda]\nonumber\\
&&-8i\theta\theta(\bar{\theta}\bar{\sigma}^l\sigma^m\mathscr{D}_l\mathscr{D}_m\bar{\lambda})+8i\theta\theta[\bar{\theta}\bar{\lambda}, d]\nonumber\\
&&+4i \theta\theta(\bar{\theta}\bar{\sigma}^k\sigma^m\partial_k \mathscr{D}_m \bar{\lambda})+4i\bar{\theta}\bar{\theta}(\theta\sigma^k\bar{\sigma}^m\partial_k \mathscr{D}_m\lambda)\nonumber\\
&&+\theta\theta\bar{\theta}\bar{\theta}\bigl(2\Box d+2i\partial^m[v_m, d]+i\mathrm{Tr}[\bar{\sigma}^m\sigma^l\bar{\sigma}^n\sigma^k]\partial_n\mathscr{D}_l v_{mk}\nonumber\\
&&-2i\partial_m\{\bar{\lambda}_{\dot{\alpha}}, (\bar{\sigma}^m\lambda)^{\dot{\alpha}}\}\bigr).\nonumber\\
\end{eqnarray}
On the other hand, $F^{\dagger}$ is represented in $(x, \theta, \bar{\theta})$ coordinates by
\begin{eqnarray}
F^{\dagger}(x, \theta, \bar{\theta})&=&-8d+8\theta\sigma^m\mathscr{D}_m\bar{\lambda}-8\bar{\theta}\bar{\sigma}^m\mathscr{D}_m\lambda\nonumber\\
&&-4(\bar{\theta}\bar{\sigma}^m\theta)[v_m, d]+4(\theta\sigma^l\bar{\sigma}^m\sigma^k\bar{\theta})\mathscr{D}_l v_{mk}+8[\bar{\theta}\bar{\lambda}, \theta\lambda]\nonumber\\
&&-4i\theta\theta(\bar{\theta}\bar{\sigma}^k\sigma^m\partial_k \mathscr{D}_m\bar{\lambda})-4i \bar{\theta}\bar{\theta}(\theta\sigma^k\bar{\sigma}^m\partial_k \mathscr{D}_m \lambda)\nonumber\\
&&+8i\bar{\theta}\bar{\theta}(\theta\sigma^l\bar{\sigma}^m\mathscr{D}_l\mathscr{D}_m\lambda)+8i\bar{\theta}\bar{\theta}[\theta\lambda, d] \nonumber\\
&&+\theta\theta\bar{\theta}\bar{\theta}\bigl(2\Box d+2i\partial^m[v_m, d]-i\mathrm{Tr}[\bar{\sigma}^m{\sigma}^k\bar{\sigma}^n{\sigma}^l]\partial_n\mathscr{D}_l v_{mk}\nonumber\\
&&+2i\partial_m\{(\bar{\lambda}\bar{\sigma}^m)^{\alpha}, \lambda_{\alpha} \}\bigr).\nonumber\\
\end{eqnarray}
Finally, we get the gauge covariant term in $(x, \theta, \bar{\theta})$ coordinates as follows.
\begin{eqnarray}
&&\hspace{0cm}\left( \frac{L_\cv}{1-e^{-L_\cv}}\cdot \left(D^{\alpha}w_{\alpha}+\{e^{-\cv}D^{\alpha}e^{\cv}, w_\alpha\}\right) \right)+~h.c.\nonumber\\
&=&-16d+16\theta\sigma^m\mathscr{D}_m\bar{\lambda}-16\bar{\theta}\bar{\sigma}^m\mathscr{D}_m\lambda\nonumber\\
&&+16\theta\sigma^m\bar{\theta}\mathscr{D}^k v_{mk}+16[\bar{\theta}\bar{\lambda}, \theta\lambda]\nonumber\\
&&-8i\theta\theta(\bar{\theta}\bar{\sigma}^l\sigma^m\mathscr{D}_l\mathscr{D}_m\bar{\lambda})+8i\theta\theta[\bar{\theta}\bar{\lambda}, d]\nonumber\\
&&+8i\bar{\theta}\bar{\theta}(\theta\sigma^l\bar{\sigma}^m\mathscr{D}_l\mathscr{D}_m\lambda)+8i\bar{\theta}\bar{\theta}[\theta\lambda, d]\nonumber\\
&&+\theta\theta\bar{\theta}\bar{\theta}\bigl(4\Box d+4i\partial^m[v_m, d]\nonumber\\
&&+i\mathrm{Tr}[\bar{\sigma}^m\sigma^l\bar{\sigma}^n\sigma^k-\bar{\sigma}^m\sigma^k\bar{\sigma}^n\sigma^l]\mathscr{D}_n\mathscr{D}_l v_{mk}\nonumber\\
&&-2i\partial_m\{\bar{\lambda}_{\dot{\alpha}}, (\bar{\sigma}^m\lambda)^{\dot{\alpha}}\}+2i\partial_m\{(\bar{\lambda}\bar{\sigma}^m)^{\alpha}, \lambda_{\alpha}\}\nonumber\\
&&-\frac{4}{3}[v_m, [v^m, d]]\bigr)
\label{mainterm}.
\end{eqnarray}
In a similar way, we obtain $\delta \cv$ in terms of $(x, \theta, \bar{\theta})$ coordinates as
%?????????????????????????????ã?L?????ã????????????????ã?ã?????$\delta V$????ã???ç???ã?ã?????????????ã?ã????????????$x, \theta, \bar{\theta}$?ã?ã???????ã?L????????ã??????
\begin{eqnarray}
\delta \cv(x, \theta, \bar{\theta})&=&\Phi_\cv+\Phi_\cv^\dagger+\frac{1}{2}[\cv, \Phi_\cv-\Phi_\cv^\dagger]+\frac{1}{12}[\cv, [\cv, \Phi_\cv+\Phi_\cv^{\dagger}]]\nonumber\\
&=&16d-16\theta\sigma^m\mathscr{D}_m\bar{\lambda}+16\bar{\theta}\bar{\sigma}^m\mathscr{D}_m\lambda-16\theta\sigma^k\bar{\theta}\mathscr{D}_k\partial_m v^m\nonumber\\
&&-8i\theta\theta\bar{\theta}\bar{\sigma}^k\sigma^m\mathscr{D}_k\mathscr{D}_m\bar{\lambda}-8\theta\theta\bar{\theta}_{\dot{\alpha}}[\bar{\lambda}^{\dot{\alpha}}, \partial_m v^m]\nonumber\\
&&-8i\bar{\theta}\bar{\theta}\theta\sigma^k\bar{\sigma}^m\mathscr{D}_k\mathscr{D}_m\lambda-8\bar{\theta}\bar{\theta}\theta^{\alpha}[\lambda_{\alpha}, \partial_m v^m]\nonumber\\
&&+4\theta\theta\bar{\theta}\bar{\theta}\bigl(\Box d+i\{\bar{\lambda}_{\dot{\alpha}}, (\bar{\sigma}^m\mathscr{D}_m \lambda)^{\dot{\alpha}}\}-i\{\lambda^{\alpha}, (\sigma^m\mathscr{D}_m \bar{\lambda})_{\alpha}\}\nonumber\\
&&+i[d, \partial_m v^m]+i[v^m, \partial_md]-\frac{1}{6}[v_m, [v^m, d] ]\bigr).\label{alphaterm}
\end{eqnarray}

\subsection{Gradient Flow Equation of Super Yang-Mills Theory for Each Component of Vector Multiplet } 
%We substitute \eref{mainterm} and \eref{alphaterm} into \eref{general1} under the WZ gauge.
Because a physical quantity does not depend on the form of the $\alpha_0$ term, we choose a particular value $\alpha_0=1$. Then we obtain 
\begin{eqnarray}
&&\hspace{0cm}\left( \frac{L_\cv}{1-e^{-L_\cv} }\cdot \left(D^{\alpha}w_{\alpha}+\{e^{-\cv}D^{\alpha}e^{\cv}, w_{\alpha}\}\right) \right)+~h.c.+1\cdot \delta \cv\nonumber\\
&=&16\theta\sigma^m\bar{\theta}\mathscr{D}^k v_{mk}+16[\bar{\theta}\bar{\lambda},\theta\lambda]-16\theta\sigma^k\bar{\theta}\mathscr{D}_k\partial_mv^m\nonumber\\
&&-16i\theta\theta\bar{\theta}\bar{\sigma}^k\sigma^m\mathscr{D}_k\mathscr{D}_m\bar{\lambda}+8i\theta\theta[\bar{\theta}\bar{\lambda}, d+i\partial_m v^m]\nonumber\\
&&+16i\bar{\theta}\bar{\theta}\theta\sigma^k\bar{\sigma}^m\mathscr{D}_k\mathscr{D}_m\lambda+8i\bar{\theta}\bar{\theta}[\theta\lambda, d-i\partial_m v^m]\nonumber\\
&&+\theta\theta\bar{\theta}\bar{\theta}\bigl(8\Box d+8i[v_m, \partial^m d]+i\mathrm{Tr}[\bar{\sigma}^m\sigma^l\bar{\sigma}^n\sigma^k-\bar{\sigma}^m\sigma^k\bar{\sigma}^n\sigma^l]\mathscr{D}_n\mathscr{D}_l v_{mk}\nonumber\\
&&+4i\{\bar{\lambda}_{\dot{\alpha}}, (\bar{\sigma}^m\mathscr{D}_m\lambda)^{\dot{\alpha}}\}-4i\{\lambda^{\alpha}, (\sigma^m\mathscr{D}_m\bar{\lambda})_{\alpha}\}-2[v_m, [v^m, d]]  \bigr).
\end{eqnarray}
Finally, we obtain the flow equations for the each component of the vector multiplet as
{\allowdisplaybreaks\begin{eqnarray}
\dot{c}&=&0,\\
\dot{\chi}&=&0,\\
\dot{\bar{\chi}}&=&0,\\
\dot{m}&=&0,\\
\dot{m}^*&=&0,\\
\dot{v}_m&=&-16\mathscr{D}^k v_{mk}+16\mathscr{D}_m\partial_k v^k-8\{\bar{\lambda}_{\dot{\alpha}},(\bar{\sigma}_m \lambda)^{\dot{\alpha}} \},\\
\dot{\bar{\lambda}}&=&-16\bar{\sigma}^k\sigma^m\mathscr{D}_k\mathscr{D}_m\bar{\lambda}+8[\bar{\lambda}, d+i\partial_m v^m],\label{matter1}\\
\dot{\lambda}&=&-16\sigma^k\bar{\sigma}^m\mathscr{D}_k\mathscr{D}_m\lambda-8[\lambda, d-i\partial_m v^m],\label{matter2}\\
\dot{d}&=&16\Box d+16i[v_m, \partial^m d]\nonumber\\
&&+2i\mathrm{Tr}[\bar{\sigma}^m\sigma^l\bar{\sigma}^n\sigma^k-\bar{\sigma}^m\sigma^k\bar{\sigma}^n\sigma^l]\mathscr{D}_n\mathscr{D}_l v_{mk}\nonumber\\
&&+8i\{\bar{\lambda}_{\dot{\alpha}}, (\bar{\sigma}^m\mathscr{D}_m\lambda)^{\dot{\alpha}}\}-8i\{\lambda^{\alpha}, (\sigma^m\mathscr{D}_m\bar{\lambda})_{\alpha}\}\nonumber\\
&&-4[v_m, [v^m, d]].   
\end{eqnarray}}
We find that the flow equations for each component are consistent with WZ gauge. Here we choose initial conditions to satisfy the WZ gauge at $t=0$ as
{\allowdisplaybreaks\begin{eqnarray}
{c}|_{t=0}&=&0,\\
{\chi}|_{t=0}&=&0,\\
{\bar{\chi}}|_{t=0}&=&0,\\
{m}|_{t=0}&=&0,\\
{m}^*|_{t=0}&=&0,\\
{v}_m|_{t=0}&=&V_m,\\
{\bar{\lambda}}|_{t=0}&=&\bar{\Lambda},\\
{\lambda}|_{t=0}&=&\Lambda,\\
{d}|_{t=0}&=&D.
\end{eqnarray}}
Let us compare the flow equation for Yang-Mills theory proposed by Luscher with our results for super Yang-Mills theory in Eqs~\eref{matter1} and \eref{matter2}. In Ref.~\cite{Luscher:2013cpa}, Luscher claims that the gradient flow equations of the quark field are given as
\begin{eqnarray}
\dot{\bar{\chi}}&=&\bar{\chi}\overleftarrow{\Delta}+\alpha_0\bar{\chi}\partial_\nu B_\nu\label{LS1},\\
\dot{\chi}&=&\Delta\chi-\alpha_0\partial_\nu B_\nu\chi\label{LS2}.
\end{eqnarray}
On the other hand our results for the gradient flow equations of the gaugino field in Eqs.~\eref{matter1} and \eref{matter2} are given as
\begin{eqnarray}
\dot{\bar{\lambda}}&=&-16\bar{\sigma}^k\sigma^m\mathscr{D}_k\mathscr{D}_m\bar{\lambda}+8[\bar{\lambda}, d+i\partial_m v^m],\label{matter11}\\
\dot{\lambda}&=&-16\sigma^k\bar{\sigma}^m\mathscr{D}_k\mathscr{D}_m\lambda-8[\lambda, d-i\partial_m v^m].\label{matter22}
\end{eqnarray}
If we regard $\Delta$ as $\ooalign{\hfil/\hfil\crcr D}^2$, Eqs.~\eref{LS1} and \eref{LS2} are almost similar to our results Eqs.~\eref{matter11} and \eref{matter22} respectively except for $[\bar{\lambda}, d]$ term and $[\lambda, d]$ term and the point that $\alpha_0$ terms are described in terms of commutation relations. 
%%%%%%%%%%%%%%%%%%%%%%%%%%%%%%%%%%%%%%%%%%%%%%%%%

\section{Summary and Discussion}\label{summary3}\label{s5}\label{allsummary}

In this paper, we proposed the generalized gradient flow equation for field theories with nonlinearly realized symmetry.  
Introducing the invariant norm for the variation of the field $\phi^a(x)$ where $a=1,\cdots, M$ is the index for the internal degrees of freedom, 
one can naturally define a metric $g_{ab}(\phi(x))$ in the functional space. Using this metric, we proposed the generalized 
gradient flow equation as
\begin{eqnarray}
\dot{\phi^a_t}(x) = - g^{ab}(\phi_t(x)) \frac{\delta S(\phi_t)}{\phi_t^b(x)}.
\end{eqnarray}
Applying  the generalized equation to super Yang-Mills theory using the superfield formalism, we obtained a gradient flow equation 
 which manifestly preserves both super symmetry and super gauge symmetry. 
 %Since in the general gauge the equation involves infinite number of terms, it is difficult to solve it non-perturbatively.  In order to obtain the gradient flow equation with finite number of terms, we need to take the WZ gauge. 
By choosing an appropriate $\alpha_0$ term described in terms of $\Phi_\cv$ in Eq.~\eref{important}, we obtained a 
gradient flow equation of the super Yang-Mills theory which is closed under the WZ gauge.

%Our result also gives a natural derivation of the gradient flow equation of the matter field at least in the adjoint matter field in super Yang-Mills theory, which could be a hint for deriving the gradient flow equation of matter fields. 

We found that the gradient flow of the super Yang-Mills theory is very similar to  the one in Yang-Mills theory 
and QCD. It is known that the gradient flow equation of the Yang-Mills theory and QCD has a wide variety of successful applications. 
We expect that our method may also be useful for testing the validity of various proposals of supersymmetric lattice models 
as well as extracting the physics of the super Yang-Mills theory.
It is important to examine whether gauge invariant physical quantities require additional renormalization or not, which is under way. 
It is also interesting to study the properties of the generalized gradient flow equation for the nonlinear sigma model, which is a subject for future studies.% They are future's work.

% \textcolor{red}{\it application for Ward-Takahashi identity, future prospects extended supersymmetry}
 
 % The problem whether the equation is also SUSY invariant or not remains to be solved. 

%%%%%%%%%%%%%%%%%%%%%%%%%%%%%%%%%%%
\section*{Acknowledgments}
We would like to thank Masanori Hanada for useful advice. We also thank Satoshi Yamaguchi,  Koji Hashimoto and 
Akinori Tanaka for fruitful discussions and comments. This work was supported by Grant-in-Aid for JSPS Fellows Grant Number
25$\cdot$1336 and Grant-in-Aid for Scientific Research (c) Grant Number 26400248.

%%%%%%%%%%%%%%%%%%%%%%%%%%%%%%%%%%5
\newpage
\appendix

\section{Notation}\label{appz}
We use the following notation.
The definition of the covariant derivative and the gauge field strength are
\begin{eqnarray}
\mathscr{D}_m\cdot&\equiv&\partial_m\cdot+\frac{i}{2}[v_m, \cdot],\\
v_{mn}&\equiv&\partial_m v_n-\partial_n v_m+\frac{i}{2}[v_m, v_n].
\end{eqnarray}
respectively. The differential operators $D$ and $\bar{D}$ are
\begin{eqnarray}
D_\alpha(x)&=&\frac{\partial}{\partial \theta^{\alpha}}+i(\sigma^m\bar{\theta})_{\alpha}\partial_m,\\
\bar{D}_{\dot{\alpha}}(x)&=&-\frac{\partial}{\partial \bar{\theta}^{\dot{\alpha}}}-i(\theta\bar{\sigma}^m)_{\dot{\alpha}}\partial_m,
\end{eqnarray}
respectively. We introduce $y$ and $y^{\dagger}$ as 
\begin{eqnarray}
y^m&=&x^m+i\theta\sigma^m\bar{\theta},\\
y^{\dagger m}&=&x^m-i\theta\sigma^m\bar{\theta}.
\end{eqnarray}
respectively. For the sake of ease, we give  $D$ and $\bar{D}$  in terms of $(y, \theta, \bar{\theta})$ or  $(y^\dagger, \theta, \bar{\theta})$ coordinates as

\begin{eqnarray}
D_\alpha(y, \theta, \bar{\theta})&=&\frac{\partial}{\partial \theta^{\alpha}}+2i(\sigma^m\bar{\theta})_{\alpha}\frac{\partial}{\partial y^m},\label{DDD}\\
\bar{D}_{\dot{\alpha}}(y, \theta, \bar{\theta})&=&-\frac{\partial}{\partial \bar{\theta}^{\dot{\alpha}}},\\
D_\alpha(y^\dagger, \theta, \bar{\theta})&=&\frac{\partial}{\partial \theta^{\alpha}},\\
\bar{D}_{\dot{\alpha}}(y^\dagger, \theta, \bar{\theta})&=&-\frac{\partial}{\partial \bar{\theta}^{\dot{\alpha}}}-2i(\theta\bar{\sigma}^m)_{\dot{\alpha}}\frac{\partial}{\partial y^{\dagger m}}.
\end{eqnarray}

\section{$O(N)$ Nonlinear Sigma Model}\label{NLS}

The $O(N)$ nonlinear sigma model is described by the unit $O(N)$ vector field $\phi^\alpha(x)$ 
($\alpha=1,\cdots,  N$)  with the constraint $\displaystyle{\sum_{\alpha=1}^N (\phi^\alpha)^2=1}$.
The action is given as 
\begin{eqnarray}
S=\frac{1}{2\lambda} \int d^Dx \sum_{\alpha=1}^N \partial_\mu \phi^\alpha \partial_\mu\phi^\alpha,
\end{eqnarray}
where $\lambda$ is the coupling. The action is invariant under the global $O(N)$ rotation
\begin{eqnarray}
\phi^\alpha(x) \rightarrow \phi^\alpha(x) + \sum_{\beta=1}^N \omega^{\alpha\beta} \phi^\beta
\end{eqnarray}
where $\omega^{\alpha\beta}$ is an infinitesimal antisymmetric tensor.  
The invariant norm  is given as 
\begin{eqnarray}
||\delta \phi||^2 = \int d^Dx \sum_{\alpha=1}^{N} (\delta \phi^\alpha(x) )^2 .
\label{eq:norm}
\end{eqnarray}
The functional space can be parameterized by  $\phi^a(x) (a=1,\cdots, N-1)$ as independent fields. 
Solving the constraint, the $N$-th component is expressed as 
\begin{eqnarray}
\phi^N(x) = \pm [1-\sum_{a=1}^{N-1} (\phi^a(x))^2]^{1/2}
\label{eq:solution}
\end{eqnarray}
Substituting Eq.(\ref{eq:solution}) into Eq.(\ref{eq:norm}), we obtain
\begin{eqnarray}
||\delta \phi||^2 = \int d^Dx \sum_{a,b=1}^{N-1}g_{ab}(\phi(x)) \delta \phi^a(x) \delta \phi^b(x), 
\end{eqnarray}
where the metric in the functional space $g_{ab}(\phi(x))$ is given by
\begin{eqnarray}
g_{ab}(\phi(x)) = \delta_{ab} + \frac{\phi^a(x) \phi^b(x)}{1-\displaystyle{\sum_{c=1}^{N-1} (\phi^c(x))^2}}.
\end{eqnarray}
In this parameterization of the functional space, the $O(N)$ symmetry is nonlinearly realized as 
\begin{eqnarray} 
\phi^a(x) \rightarrow \phi^{\prime a}(x) &=&  \phi^a(x) +\delta \phi^a(x) \nonumber\\
&= &
\phi^a(x) + \sum_{b=1}^{N-1} \omega^{ab} \phi^b(x) \pm  \omega^{aN}  [1-\sum_{b=1}^{N-1} (\phi^b(x))^2]^{1/2},
\nonumber\\
\label{eq:ON}
\end{eqnarray} 
where $\omega^{ab}, \omega^{aN}$ are the infinitesimal parameters for the $O(N)$ rotation.
One can easily find that if one considers the transformation in Eq.(\ref{eq:ON})  as the coordinate transformation 
of the functional space, it is the isometry. In other words, 
\begin{eqnarray}
g^{ab}(\phi^\prime) 
%\end{eqnarray}
%holds, where $g^{\prime ab}(\phi^\prime)$ is defined as 
%\begin{eqnarray}
%g^{\prime ab}(\phi^\prime) 
= \frac{\partial \phi^{\prime a}}{\partial \phi^c}
  \frac{\partial \phi^{\prime b}}{\partial \phi^d}
  g^{cd}(\phi)
\end{eqnarray}
holds.
It is then obvious that the following generalized equation
\begin{eqnarray}
\dot{\phi}_t^a(x) = -g^{ab}(\phi_t(x)) \frac{\delta S(\phi_t)}{\delta \phi_t^b(x)}
\label{eq:GGFlow}
\end{eqnarray}
gives essentially identical time evolution for $\phi$ and $\phi^\prime$.
Moreover, a straightforward calculation shows that the gradient flow equation based on our proposal
Eq.(\ref{eq:GGFlow}) 
gives 
\begin{eqnarray}
\dot{\phi}_t^a(x) 
& =\displaystyle{ \frac{1}{\lambda}\left[
\square \phi_t^a(x) - \phi_t^a(x)
\left( \sum_{b=1}^{N-1} \phi_t^b(x) \square \phi_t^b(x) 
+ \phi_t^N(x) \square \phi_t^N(x) \right) \right]}, ~~~~
\label{eq:GFlow_ONa}
\end{eqnarray}
with $\displaystyle{\phi_t^N(x) \equiv \pm \left[1-\sum_{c=1}^{N-1}(\phi_t^c(x))^2)\right]^{1/2}}$. 
The time evolution for $\phi_{t}^{N}(x)$ can also be induced using Eq.(\ref{eq:GFlow_ONa}) which 
reads
\begin{eqnarray}
\dot{\phi}_t^N(x) 
& =& \frac{1}{\lambda}\left[
\square \phi_t^N(x) - \phi_t^N(x)
\left( \sum_{b=1}^{N-1} \phi_t^b(x) \square \phi_t^b(x) 
+ \phi_t^N(x) \square \phi_t^N(x) 
\right)\right].
\nonumber\\
\label{eq:GFlow_ONN}
\end{eqnarray}
Eqs.(\ref{eq:GFlow_ONa}), (\ref{eq:GFlow_ONN}) can be combined to
\begin{eqnarray}
\dot{\phi}^\alpha(x) 
& =& \frac{1}{\lambda}\left[
\square \phi^\alpha(x) - \phi^\alpha(x)
\left( \sum_{\beta=1}^N \phi^\beta(x) \square \phi^\beta(x) \right)\right], 
\end{eqnarray}
which is manifestly $O(N)$ symmetric and also keeps the constraint 
$\displaystyle{\sum_{\alpha=1}^N(\phi^\alpha(x))^2=1}$.

\section{Lattice Gauge Theory}\label{LGT}

%\subsubsection{$SU(N)$ lattice gauge theory}
%The $SU(N)$ lattice gauge theory is defined  on the lattice 
%with the action given as 
%\begin{eqnarray}
%S= \beta \sum_x \left[ 1 - \frac{1}{N}\mathrm{Tr} (P_{\mu,\nu}(x) + P_{\mu,\nu}^\dagger(x) )
%\right].
%\end{eqnarray}
%Here  $U_\mu(x)$ are $N\times N$ $SU(N)$ matrices on the link emenating from the lattice site $x$ towards $\mu$ direction 
%and  $P_{\mu, \nu}(x)$ is deifined as
%\begin{eqnarray}
%P_{\mu, \nu}(x)= U_\mu(x) U_\nu(x+\hat{\mu}) U^\dagger_\mu(x+\hat{\nu}) U^\dagger_\nu(x).
%\end{eqnarray}
%This action is invariant under the gauge transformation 
%\begin{eqnarray}
%U_\mu(x) \rightarrow \Lambda(x) U_\mu(x) \Lambda^\dagger(x+\hat{\mu}) .
%\end{eqnarray}
%Using the fact that the invariant norm is given as 
%\begin{eqnarray}
%||\delta U_\mu||^2 = \sum_x \mathrm{Tr}\left[\delta U_\mu^\dagger(x)\delta U_\mu(x)
%\right],
%\end{eqnarray}
%where $\Lambda(x)$ are arbitrary $SU(N)$ matrices on the lattice site $x$.
% One can show that the generalized gradient flow equation is given as 
%\begin{eqnarray}
% \dot{U}_t(\mu,x) =\beta \sum_{\nu(\neq\mu)}
%\left[ F_t(\mu,\nu,x) -\frac{1}{N} \mbox{tr}( F_t(\mu,\nu,x) )\right]
% U_t(\mu,x),
% \label{eq:lattice}
%\end{eqnarray}
%where
%\begin{eqnarray}
%F_t(\mu,\nu,x)
%\equiv  \left[ P_t(\mu, \nu,x) - P^\dagger_t(\mu,\nu,x) 
% +  P_t(\mu, -\nu, x) - P^\dagger(\mu, -\nu,x)  \right] 
%\end{eqnarray}

%See Appendix \ref{LGT} for derivation of Eq. (\ref{eq:lattice}).

The $SU(N)$ lattice gauge theory is described by the link variable 
$U(\mu,x)$ which are $N\times N$ $SU(N)$ matrices . The action  is given as 
\begin{eqnarray}
S= \beta \sum_x \sum_{\mu> \nu}\left[ 1 - \frac{1}{N}\mathrm{Tr} (P(\mu,\nu,x) + P^\dagger(\mu,\nu,x) )
\right].
\end{eqnarray}
Here  $P(\mu, \nu,x)$ is the plaquette defined as
\begin{eqnarray}
P(\mu, \nu,x)= U(\mu, x) U(\nu, x+\hat{\mu}) U^\dagger(\mu,x+\hat{\nu}) U^\dagger(\nu,x).
\end{eqnarray}
This action is invariant under the gauge transformation 
\begin{eqnarray}
U(\mu,x) \rightarrow \Lambda(x) U(\mu,x) \Lambda^\dagger(x+\hat{\mu}) , 
\end{eqnarray}
where $\Lambda(x)$ are arbitrary $SU(N)$ matrices on the lattice site $x$
and the invariant norm is given as 
\begin{eqnarray}
||\delta U||^2 = \sum_x \sum_\mu \mathrm{Tr}\left[\delta U^\dagger(\mu, x)\delta U(\mu,x)
\right].
\end{eqnarray}
The link variable $U(\mu,x)$ can be parameterized as 
\begin{eqnarray}
U(\mu,x) = \exp(i A_\mu(x)), 
\end{eqnarray}
where $\displaystyle{A_\mu(x)\equiv \sum_{a=1}^{N^2-1}A_\mu^a(x) T^a}$  is  $SU(N)$ gauge field  and $T^a (a=1,\cdots N^2-1)$ are traceless 
Hermitian $N\times N$ matrices with the condition $\mbox{Tr}(T^a T^b)=\frac{1}{2} \delta^{ab}$.

The metric from the invariant norm can be explicitly obtained using the following matrix identity.
Let $V$ be a $N\times N$ matrix and consider an infinitesimal variation $\delta V$.  
Defining an linear operator $L_V$ which acts on arbitrary matrix $M$ as 
\begin{eqnarray}
L_V \cdot M \equiv \left[V,M\right],
\end{eqnarray}
then the following matrix identity holds for linear order in $\delta V$
\begin{eqnarray}
e^{-V}(e^{V+\delta V} - e^V)  = \frac{1-e^{-L_V}}{L_V}\cdot \delta V .
\end{eqnarray}
Using this matrix identity and setting $V= i A_\mu(x)$, the invariant norm can be rewritten as
\begin{eqnarray}
||\delta U||^2 =  \sum_{x,\mu }\mathrm{Tr}
\left[
\left(\frac{1-e^{-L_V}}{L_V}\cdot T^a\right) 
\left(\frac{1-e^{-L_V}}{L_V}\cdot T^b\right) 
\right]
\delta A^a_\mu(x) \delta A^b_\mu(x) .
\end{eqnarray}
Therefore, the metric $g_{ab}(A_\mu(x))$ becomes
\begin{eqnarray}
g_{ab}(A_\mu(x)) = \mbox{Tr} 
\left[
\left(\frac{1-e^{-L_V}}{L_V}\cdot T^a\right) 
\left(\frac{1-e^{-L_V}}{L_V}\cdot T^b\right) 
\right]. 
\end{eqnarray}
A simple algebra shows that the metric $g^{ab}(A_\mu(x))$ is 
\begin{eqnarray}
g^{ab}(A_\mu(x)) = 4 \mbox{Tr} 
\left[
\left(\frac{L_V}{1-e^{-L_V}} \cdot T^a\right) 
\left(\frac{L_V}{1-e^{-L_V}} \cdot T^b\right) 
\right]. 
\end{eqnarray}
The generalized gradient flow equation for the field $A_{t \mu}(x)$
\begin{eqnarray}
\dot{A}^a_{t \mu}(x) = - g^{ab}(A_\mu(x)) \frac{\delta S(A_t)}{\delta A^b_{t \mu}(x)}
\end{eqnarray}
gives 
\begin{eqnarray}
\dot{A}_{t \mu}(x) &=& -i \beta  \frac{L_V}{1-e^{-L_V}} 
\nonumber\\
&&\cdot
 \left(
X_t(\mu,x) U_t(\mu,x) - U^\dagger_t(\mu,x) X^\dagger_t(\mu,x)
\right.
\nonumber\\
& &
\left.
-\frac{1}{N}\mbox{tr}
\left[ X_t(\mu,x) U_t(\mu,x) - U^\dagger_t(\mu,x) X^\dagger_t(\mu,x)
\right]
\right)
\label{eq:GFlow_A}
\end{eqnarray}
where $V_t, U_t(\mu,x), X_t(\mu,x)$ are defined as 
\begin{eqnarray}
V_t  &=& i A_{t\mu}(x)\\
U_t(\mu,x) &=&\exp(iA_{t\mu}(x))\\
X_t(\mu,x) 
&=& \sum_{\nu \neq\mu}  
\left[ U_t(\nu,x+\hat{\mu} )U_t^\dagger(\mu,x+\hat{\nu} )U_t^\dagger(\nu,x) \right.\nonumber\\
&&  \left.    - U_t^\dagger(\nu,x+\hat{\mu}-\hat{\nu}) U_t^\dagger(\mu,x-\hat{\mu}) U(\nu,x-\hat{\nu} )\right]
\end{eqnarray}
Using Eq.(\ref{eq:GFlow_A}), we obtain the generalized gradient flow equation for $U_t(\mu,x)$ as 
\begin{eqnarray}
&&\dot{U}_t(\mu,x)
\nonumber\\
%&=& i \frac{1-e^{-L_V}}{L_V} 
%\cdot  \dot{A}_{t \mu}(x) \\
&=&  \beta  
\left(
U_t(\mu,x) X_t(\mu,x) - X^\dagger_t(\mu,x)U^\dagger_t(\mu,x)
\right. 
\nonumber\\
&&
\left.
-\frac{1}{N} \mbox{tr}\left[X_t(\mu,x) U_t(\mu,x) - U^\dagger_t(\mu,x) X^\dagger_t(\mu,x)\right]
\right) U_t(\mu,x)
\end{eqnarray}
Noting that 
\begin{eqnarray}
U_t(\mu,x) X_t(\mu,x)
= \sum_{\nu\neq\mu}
P_t(\mu,\nu,x)  +P_t(\mu,-\nu,x) 
\end{eqnarray}
where $P_t(\mu,\nu,x)$ is the plaquette constructed from $U_t$, the final form for the generalized gradient flow equation 
for the link field in $SU(N)$ lattice gauge theory becomes
\begin{eqnarray}
&&\dot{U}_t(\mu,x)
\nonumber\\
&=&  \beta  
\sum_{\nu\neq\mu} 
\left(
P_t(\mu,\nu,x)+ P_t(\mu,-\nu,x)- P^\dagger_t(\mu,\nu,x) - P^\dagger_t(\mu,-\nu,x)
\right. 
\nonumber\\
&&
\left.
-\frac{1}{N} \mbox{Tr}
(P_t(\mu,\nu,x)+ P_t(\mu,-\nu,x)- P^\dagger_t(\mu,\nu,x) - P^\dagger_t(\mu,-\nu,x))
\right) U_t(\mu,x)
\nonumber\\
\end{eqnarray}
which agrees with Eq.(1.4)  given in the paper \cite{Luscher:2010iy}.
%%%%%%%%%%%%%%%%%%%%%%%%%%%%%%%%%%%%%%%%%%%%
%%%%%%%%%%%%%%%%%%%%%

\section{Short Summary of Supersymmetry}\label{summaryofsusy}
We give the notation of the superfield formalism. We follow the convention by Wess and Bagger \cite{wess}.
\subsection{Definition}
The chiral superfield is defined by 
\begin{eqnarray}
\bar{D}_{\alpha}\Phi=0.
\end{eqnarray}
We described chiral multiplet  $\Phi=\{A, \psi, F \}$ in terms of $(x, \theta, \bar{\theta})$ coordinates as 
\begin{eqnarray}
\Phi(x, \theta, \bar{\theta})&=&A+i\theta\sigma^m\bar{\theta}\partial_m A+\frac{1}{4}\theta\theta\bar{\theta}\bar{\theta}\Box A\nonumber\\
&&\sqrt{2}\theta\psi-\frac{i}{\sqrt{2}}\theta\theta\partial_m\psi\sigma^m\bar{\theta}+\theta\theta F
\end{eqnarray}
The vector superfield is defined by
\begin{eqnarray}
V=V^{\dagger}.
\end{eqnarray}
We described vector multiplet $V=\{C,X,\bar{X},M,M^*,V_m,\Lambda, \bar{\Lambda},D\}$ in terms of $(x, \theta, \bar{\theta})$ coordinates as 
\begin{eqnarray}
V(x, \theta, \bar{\theta})&=&C+i\theta X -i\bar{\theta}\bar{X}+\frac{i}{2}\theta\theta M-\frac{i}{2}\bar{\theta}\bar{\theta} M^*\nonumber\\
&&-\theta\sigma^m \bar{\theta}V_m+i\theta\theta\bar{\theta}[\bar{\Lambda}+\frac{i}{2}\bar{\sigma}^m\partial_m X]\nonumber\\
&&-i\bar{\theta}\bar{\theta}\theta[\Lambda+\frac{i}{2}\sigma^m\partial_m\bar{X}]+\frac{1}{2}\theta\theta\bar{\theta}\bar{\theta}[D+\frac{1}{2}\Box C].
\end{eqnarray}
\subsection{Wess-Zumino Gauge}
The infinitesimal super gauge transformation is defined by
\begin{eqnarray}
V'=V+\Phi+\Phi^{\dagger}.
\end{eqnarray}
Under this transformation, the each component of the vector multiplet transforms as follows:
\begin{eqnarray}
C'&=&C+A+A^*\\
X'&=&X-i\sqrt{2}\psi\\
M'&=&M-2iF\\
V'_m&=&V_m-i\partial_m(A-A^*)\\
\Lambda'&=&\Lambda\\
D'&=&D
\end{eqnarray}
Using this gauge transformation, we fixed the WZ gauge, which is $C, X, M=0$. Under this gauge, V is described in terms $(x, \theta, \bar{\theta})$ coordinates as
\begin{eqnarray}
V(x, \theta, \bar{\theta})&=&-\theta\sigma^m\bar{\theta}V_m+i\theta\theta\bar{\theta}\bar{\Lambda}-i\bar{\theta}\bar{\theta}\Lambda+\frac{1}{2}\theta\theta\bar{\theta}\bar{\theta}D,\\
V^2(x, \theta, \bar{\theta})&=&-\frac{1}{2}\theta\theta\bar{\theta}\bar{\theta}V_mV^m,\\
V^3(x, \theta, \bar{\theta})&=&0.
\end{eqnarray}
And V is also described in terms $(y, \theta, \bar{\theta})$ coordinate as
\begin{eqnarray}
V(y, \theta, \bar{\theta})&=&-\theta\sigma^m\bar{\theta}V_m+i\theta\theta\bar{\theta}\bar{\Lambda}-i\bar{\theta}\bar{\theta}\theta\Lambda\nonumber\\
&&+\frac{1}{2}\theta\theta\bar{\theta}\bar{\theta}[D-i\partial_m V^m],\\
V^2(y, \theta, \bar{\theta})&=&-\frac{1}{2}\theta\theta\bar{\theta}\bar{\theta}V_m V^m, \\
V^3(y, \theta, \bar{\theta})&=&0.
\end{eqnarray}

%%%%%%%%%%%%%%%%%%%%%%%%%%%%%%%%%%%%%%%%%%%%%%%%%%%%%%%
\section{Derivation of Gradient Flow Equation for Vector Superfield $\cv$ }\label{metric}\label{explicit}
%%%%%%%%%%%%%%%%%%%%%%%%%%%%%%%%%
The invariant norm for the variation of the vector superfield $V$ is given as
\begin{eqnarray}
||\delta V||^2&=& - \int d^8 z \mathrm{Tr}[e^{-V}\delta e^V e^{-V}\delta e^V] .
\end{eqnarray}
The superfield $V$ can be expanded as
\begin{eqnarray}
V= V^a T^a,
\end{eqnarray}
where $T^a (a=1,\cdots, N^2-1)$ are the basis of $N\times N$ traceless Hermitian matrices with the condition
$\mathrm{Tr}(T^aT^b) = \frac{1}{2} \delta^{ab}$. 
Using the matrix identity for infinitesimal variation $\delta V$
\begin{eqnarray}
e^{-V}(e^{V+\delta V}-e^V) = \frac{1-e^{-L_V}}{L_V}\cdot \delta V,
\end{eqnarray}
one can rewrite the invariant norm as follows:
\begin{eqnarray}
||\delta V||^2&=&-\int d^8 z \delta V^a(z)\delta V^b(z)
\mathrm{Tr}\left[\left(\frac{1-e^{-L_V}}{L_V}\cdot T^a\right)\left(\frac{1-e^{-L_V}}{L_V}\cdot T^b\right)\right].~~~~~~~
%&=&\int d^8 z \delta V^a (z) \delta V^b (z) \mathrm{Tr}\left[
%g^{ab}(V)=\mathrm{Tr}\left[\left(\frac{1-e^{-L_V}}{L_V}T^a\right)\left(\frac{1-e^{-L_V}}{L_V}T^b\right)\right].\label{met}
\end{eqnarray}
Thus the metric $g_{ab}(V)$ is defined as
\begin{eqnarray}
g_{ab}(V)=-\mathrm{Tr}\left[\left(\frac{1-e^{-L_V}}{L_V}\cdot T^a\right)\left(\frac{1-e^{-L_V}}{L_V}\cdot T^b\right)\right].\label{met}
\end{eqnarray}
The metric $g^{ab}(V)$, which is the inverse of the above is then defined as
\begin{eqnarray}
g^{ab}(V)=-4\mathrm{Tr}\left[\left(\frac{L_V}{1-e^{-L_V}}\cdot T^a\right)\left(\frac{L_V}{1-e^{-L_V}}\cdot T^b\right)\right].\label{met2}
\end{eqnarray}
To derive $g^{ab}(V)$, we have used the matrix identity 
\begin{eqnarray}
\mathrm{Tr}(A T^a) \mathrm{Tr}(T^a B) = \frac{1}{2}\mathrm{Tr}(AB)\label{trab}
\end{eqnarray}
for arbitrary traceless matrices $A,B$.
%%%%%%%%%%%%%%%%%%%%%%%%%%%%%%%%%%%%%%%

The super Yang-Mills action is given as
\begin{eqnarray}
S_{\mathrm{SYM}}&=&-\int d^4 x\int d^2 \theta \mathrm{Tr}[W^\alpha W_\alpha]+h.c.\\
&=&\int d^8 z\mathrm{Tr}[e^{-V}(D^\alpha e^V) W_\alpha]+h.c..
\end{eqnarray}
When we make a variation over the $V^b$ field, we obtain
\begin{eqnarray}
\frac{\delta S_{\mathrm{SYM}}}{\delta V^b(z)}&=&\int d^8 w\mathrm{Tr}[\frac{\delta}{\delta V^a(z)}\{e^{-V}(D^\alpha e^V) W_\alpha\}(w)]+h.c.\\
&=&2\int d^8w \mathrm{Tr}\Bigl[\frac{\delta e^{V(w)}}{\delta V^b(z)}\{(D^\alpha W_\alpha) e^{-V}+W^\alpha (D_\alpha e^{-V})\}(w)\nonumber\\
&&\hspace{1.7cm}-\frac{\delta e^{-V(w)}}{\delta V^b(z)}(D^\alpha e^V) W_\alpha (w)   \Bigr]+h.c.~~~\\
&=&\mathrm{Tr}\left[T^b \frac{e^{L_V}-1}{L_V}\cdot  \left(D^{\alpha} W_{\alpha}+\{e^{-V}D^\alpha e^{V}, W_\alpha\}\right)(z)\right]+h.c.~~~~~~~~~~\label{gab2}
\end{eqnarray}
%where the normalization condition, 
%\begin{eqnarray}
%\mathrm{Tr}[T^a T^b]=\frac{1}{2}\delta^{ab}\end{eqnarray}
% is imposed. 
Here we used the useful formulae as
\begin{eqnarray}
\delta(e^V)&=&e^V\left[ \frac{1-e^{-L_V}}{L_V}  \cdot\delta V\right]\label{gc0}\\
&=&\left[ \frac{e^{L_V}-1}{L_V} \cdot\delta V \right] e^V,\\
\delta(e^{-V})&=&e^{-V}\left[ \frac{1-e^{L_V}}{L_V} \cdot\delta V\right] \\
&=&\left[ \frac{e^{-L_V}-1}{L_V}  \cdot\delta V \right]e^{-V}.
\end{eqnarray}
%For example, we give the proof of Eq.~\eref{gc0}. We define $\delta\chi(t)$ as
%\begin{eqnarray}
%\delta\chi(t)&\equiv& e^{-tV}e^{t(V+\delta V)}-1\label{gc1}\\
%&=&e^{-tV}\delta(e^{tV}),\label{gc4}
%\end{eqnarray}
%and we differentiate Eq.~\eref{gc1} in respect to $t$. We obtain
%\begin{eqnarray}
%\delta\dot{\chi}(t)&=-[V, \delta\chi(t)]+\delta V.\label{gc2}
%\end{eqnarray}
%The solution of Eq.~\eref{gc2} is given as
%\begin{eqnarray}
%\delta\chi(t)&=&\left( \frac{1-e^{-tL_V}}{L_V} \right) \cdot\delta V.\label{gc3}
%\end{eqnarray}
%When we set $t=1$ in Eqs~\eref{gc4} and \eref{gc2}, we obtain Eq.~\eref{gc0}.
%Other equations also can be proved in a similar way or using the formula
%\begin{eqnarray}
%e^V X e^{-V}=e^{L_V}X
%\end{eqnarray}
%where $X$ is any function.
%%%%%%%%%%%%%%%%%%%%%%%%%%%%%%%%%%%%%%
Combining Eqs.\eref{met2} and \eref{gab2}, and replacing the $V$ field with the $\cv$ field, we obtain 
\begin{eqnarray}
g^{ab}(\cv)\frac{\delta S_{\mathrm{SYM}}}{\delta \cv^b(z)}&=&-4\mathrm{Tr}\left[\left(\frac{L_\cv}{1-e^{-L_\cv}}\cdot T^a\right)\left(\frac{L_\cv}{1-e^{-L_\cv}}\cdot T^b\right)\right]\nonumber\\
&&\times \mathrm{Tr}\left[T^b \frac{e^{L_\cv}-1}{L_\cv}\cdot  \left(D^{\alpha} w_{\alpha}+\{e^{-\cv}D^\alpha e^{\cv}, w_\alpha\}\right)(z)\right]+h.c.~~~~~~~~~~~\\
&=&-2\mathrm{Tr}\left[T^a \frac{L_\cv}{1-e^{-L_\cv}}\cdot  \left(D^{\alpha} w_{\alpha}+\{e^{-\cv}D^\alpha e^{\cv}, w_\alpha\}\right)(z)\right] +h.c..~~~~~~~~\label{gabSb}
\end{eqnarray}
Here, we used the identity in Eq.\eref{trab}.
%%%%%%%%%%%%%%%%%%%

The matrix form of the gradient flow equation is 
\begin{eqnarray}
\dot{\cv}=-T^a g^{ab}\frac{\delta S_{\mathrm{SYM}}}{\delta \cv^b}+\alpha_0 \delta \cv.
\end{eqnarray} 
Using the matrix identity
\begin{eqnarray}
T^a \mathrm{Tr}\left[T^a A \right]=\frac{1}{2}A,
\end{eqnarray}
for arbitrary traceless matrix $A$ and substituting Eq.\eref{gabSb}, we finally obtain
\begin{eqnarray}
\dot{\cv}=\frac{L_\cv}{1-e^{-L_\cv}}\cdot  \left(D^{\alpha} w_{\alpha}+\{e^{-\cv}D^\alpha e^{\cv}, w_\alpha\}\right) +h.c.+\alpha_0 \delta \cv.
\end{eqnarray}
 
%%%%%%%%%%%%%%%%%%

\section{Pure Abelian Supersymmetric Theory}\label{SUSY Abelian}\label{s2}
We consider a supersymmetric pure Abelian gauge theory to simplify the discussion. Because this theory does not have an interaction, the theory also does not have divergences in the first place, but it is useful to understand the basic structure as a toy model.

\subsection{Derivation of Gradient Flow Equation of Pure Abelian Supersymmetric Theory }
From the discussion in Sec.~\ref{proposal}, we obtain the gradient flow Equation of the pure Abelian supersymmetric theory.
The free vector field action which is invariant under the supersymmetric gauge transformation is
\begin{eqnarray}
S&=&-\frac{1}{4} \int d^4x(W^{\alpha}W_{\alpha}|_{\theta\theta}+\bar{W}_{\dot{{\alpha}}}\bar{W}^{\dot{{\alpha}}}|_{\bar{\theta}\bar{\theta}})\nonumber\\
%&=&-\int d^8 z (W^{\alpha}D_{\alpha}V+\bar{W}_{\dot{\alpha}}\bar{D}^{\dot{\alpha}}V)\\
&=&-\frac{1}{4}\int d^8 z (D^{\alpha}W_{\alpha}+\bar{D}_{\dot{\alpha}}\bar{W}^{\dot{\alpha}})V
\end{eqnarray}
where $V$ is vector multiplet, $V=\{C,X,\bar{X},M,M^*,V_m,\Lambda, \bar{\Lambda},D\}$. 
$W$ and $\bar{W}$ are defined by
\begin{eqnarray}
W_{\alpha}&=&-\bar{D}\bar{D}D_{\alpha}V,\\
\bar{W}_{\dot{\alpha}}&=&-{D}{D}\bar{D}_{\dot{\alpha}}V.
\end{eqnarray}
Making variation of the action $S$ over $V$, we obtain
\begin{eqnarray}
\frac{\delta S}{\delta V}%&=&\frac{1}{2}(D^\alpha W_\alpha+\bar{D}_{\dot{\alpha}}\bar{W}^{\dot{\alpha}})\\
&=&-D^\alpha W_\alpha.
\end{eqnarray}
We used here the relation equation, 
\begin{eqnarray}
D^{\alpha}W_{\alpha}=\bar{D}_{\dot{\alpha}}\bar{W}^{\dot{\alpha}}.
\end{eqnarray}
Then we obtained the extended gradient flow equation of the pure supersymmetric theory as
\begin{eqnarray}
&&\dot{\cv}=D^{\alpha}w_{\alpha}+\alpha_0(D^2\bar{D}^2+\bar{D}^2D^2)\cv,\label{2-1}\\
&&\cv|_{t=0}=V,~~w_{\alpha}|_{t=0}=W_{\alpha}.
\end{eqnarray}
where $\cv$ is vector multiplet depending on the flow time,  $\cv=\{c,\chi,\bar{\chi},m,m^*,v_m,\lambda, \bar{\lambda},d\}$. The $\alpha_0$ term, which is the second term of the R.H.S. of Eq.~\eref{2-1}, is introduced to suppress the new gauge degrees of freedom under the evolution in the flow time. %The $\alpha_0$ term may not be unique, but we here only show that the form Eq.~\eref{2-1} is adequate. %We postpone to explain how to determine the $\alpha_0$ term to Sec.~\ref{gaugefixing}. 

\subsection{Gradient Flow Equation of Pure Yang-Mills Theory for Each Component of Vector Multiplet }

Describing the extended gradient flow equation in the coordinate of superspace which are labeled $(x, \theta, \bar{\theta})$, we find out the each dependence of the component of vector multiplet on the flow time.   
\begin{eqnarray}
\cv(x, \theta, \bar{\theta})&=&c+i\theta \chi -i\bar{\theta}\bar{\chi}+\frac{i}{2}\theta\theta m-\frac{i}{2}\bar{\theta}\bar{\theta} m^*\nonumber\\
&&-\theta\sigma^m \bar{\theta}v_m+i\theta\theta\bar{\theta}[\bar{\lambda}+\frac{i}{2}\bar{\sigma}^m\partial_m \chi]\nonumber\\
&&-i\bar{\theta}\bar{\theta}\theta[\lambda+\frac{i}{2}\sigma^m\partial_m\bar{\chi}]+\frac{1}{2}\theta\theta\bar{\theta}\bar{\theta}[d+\frac{1}{2}\Box c]\label{13}
\end{eqnarray}
Using \eref{13}, we calculate each terms of the gradient flow equation, we obtain
\begin{eqnarray}
D^{\alpha}w_{\alpha}&=&-2d+2\theta\sigma^m\partial_m\bar{\lambda}-2\bar{\theta}\bar{\sigma}^m\partial_m\lambda+2(\theta\sigma^k \bar{\theta})\partial^m v_{km}\nonumber\\
&&-i\bar{\theta}\bar{\theta}\theta\Box \lambda+i\theta\theta\bar{\theta}\Box\bar{\lambda}+\frac{1}{2}\theta\theta\bar{\theta}\bar{\theta}\Box d,\label{14-1}
\end{eqnarray}
\begin{eqnarray}
(D^2\bar{D}^2+\bar{D}^2D^2)\cv&=&16(d+\Box c)-16\theta(\sigma^m\partial_m\bar{\lambda}-i\Box \chi)+16\bar{\theta}(\bar{\sigma}^m\partial_m\lambda-i\Box\bar{\chi})\nonumber\\
&&+8i\theta\theta\Box m-8i\bar{\theta}\bar{\theta}\Box m^*
-16(\theta\sigma^m\bar{\theta})\partial_m\partial^k v_k \nonumber\\
&&+8i\theta\theta\bar{\theta}(\Box \bar{\lambda}+i\bar{\sigma}^m\partial_m\Box\chi) -8i\bar{\theta}\bar{\theta}\theta(\Box \lambda+i \sigma^m\partial_m\Box\bar{\chi})\nonumber\\
&&+ 4\theta\theta\bar{\theta}\bar{\theta}(\Box d +\Box \Box c).\label{14-2}
\end{eqnarray}
Substituting \eref{14-1} and \eref{14-2} into \eref{2-1}, finally, we obtain the flow equations for the each component of the vector multiplet as
%\begin{eqnarray}
%\dot{c}&=&-2d+16\alpha_0(d+\Box c)\\
%i\dot{\chi}&=&2\sigma^m\partial_m\bar{\lambda}-16\alpha_0(\sigma^m\partial_m\bar{\lambda}-i\Box \chi)\\
%i\dot{\bar{\chi}}&=&2\bar{\sigma}^m\partial_m\lambda-16\alpha_0(\bar{\sigma}^m\partial_m\lambda-i\Box \bar{\chi})\\
%\dot{m}&=&16\alpha_0\Box m\\
%\dot{m}^*&=&16\alpha_0\Box m^*\\
%\dot{v}_m&=&-2\partial^k v_{mk}^{(0)}+16\alpha_0\partial_m\partial^l v_l\\
%\dot{\bar{\lambda}}+\frac{i}{2}\bar{\sigma}^m\partial_m\dot{{\chi}}&=&\Box \bar{\lambda}+8\alpha_0(\Box\bar{\lambda}+i\bar{\sigma}^m\partial_m\Box \chi)\\
%\dot{\lambda}+\frac{i}{2}\sigma^m\partial_m\dot{\bar{\chi}}&=&\Box \lambda+8\alpha_0(\Box\lambda+i\sigma^m\partial_m\Box \bar{\chi})\\
%\dot{d}+\frac{1}{2}\Box\dot{c}&=&\Box d +8\alpha_0(\Box d+\Box \Box c)
%\end{eqnarray}
\begin{eqnarray}
\dot{c}&=&16\alpha_0\Box c-2(1-8\alpha_0)d,\\
\dot{\chi}&=&16\alpha_0 \Box \chi-2i (1-8\alpha_0)\sigma^m\partial_m\bar{\lambda},\\
\dot{\bar{\chi}}&=&16\alpha_0 \Box \bar{\chi}-2i (1-8\alpha_0)\bar{\sigma}^m\partial_m \lambda,\\
\dot{m}&=&16\alpha_0\Box m,\\
\dot{m}^*&=&16\alpha_0\Box m^*,\\
\dot{v}_m&=&2\Box v_m -2(1-8\alpha_0)\partial_m\partial^k v_k,\\
\dot{\bar{\lambda}}&=&2\Box \bar{\lambda},\\
\dot{\lambda}&=&2\Box \lambda,\\
\dot{d}&=&2\Box d.
\end{eqnarray}
Taking $\alpha_0$ as
{\allowdisplaybreaks\begin{eqnarray}
\alpha_0=\frac{1}{8},
\end{eqnarray}
we obtain 
\begin{eqnarray}
\dot{c}&=&2\Box c,\\
\dot{\chi}&=&2\Box \chi,\\
\dot{\bar{\chi}}&=&2\Box \bar{\chi},\\
\dot{m}&=&2\Box m,\\
\dot{m}^*&=&2\Box m^*, \\
\dot{v}_m&=&2\Box v_m, \\
\dot{\bar{\lambda}}&=&2\Box \bar{\lambda},\\
\dot{\lambda}&=&2\Box\lambda, \\
\dot{d}&=&2\Box d.
\end{eqnarray}}One can see that each component of the vector multiplet evolves separately in time.

%Looking at these equation, the each component of the vector multiplet describe separately by others, the flow does not mix them with each other. Incidentally, changing the scale of the flow time, the coefficients in front of the d'Alembert operator can be adjusted for any value.

\subsection{Flow Time Dependence of Super Gauge Transformation}
When we demand that the gradient flow equation~\eref{2-1} is invariant under the super gauge transformation, \begin{eqnarray}
\cv'=\cv+\phi+\phi^{\dagger},
\end{eqnarray}
at each time, $\phi$ have to satisfy the equation as
\begin{eqnarray}
\dot{\phi}=\alpha_0\bar{D}^2 D^2 \phi,\label{2-2}\\
\phi |_{t=0}=\Phi,
\end{eqnarray}
where $\Phi$ is a chiral field,
\begin{eqnarray}
\bar{D}\Phi=0.
\end{eqnarray}
The chirality of the $\phi$ at each flow time is guaranteed by Eq.~\eref{2-2}.

\section{Expansion of Equation \eref{Aterm} with Component Fields  }\label{E}
For the convenience of the expansion of \eref{Aterm} with the component fields, we give useful methods and formulae.
\subsection{Coordinate Transformation}
It is useful to calculate $w_\alpha$ in terms of $(y, \theta, \bar{\theta})$ coordinates. 
We obtain $w_\alpha$ as 
\begin{eqnarray}
w_\alpha(y, \theta, \bar{\theta})&=&-\bar{D}^2(e^{-\cv}D_\alpha e^\cv)\\
&=&-4i\lambda_{\alpha}+4\theta_\alpha d-2i(\sigma^m \bar{\sigma}^k \theta)_{\alpha}v_{mk}\nonumber\\
&&+4\theta\theta\{\sigma^m\mathscr{D}_m\bar{\lambda}\}_\alpha.
\end{eqnarray}
Using the expansion formula,
\begin{eqnarray}
f(y, \theta, \bar{\theta})=f(x)+i\theta\sigma^m\bar{\theta}\partial_m f(x)+\frac{1}{4}\theta\theta\bar{\theta}\bar{\theta}\Box f(x),
\end{eqnarray}
and
\begin{eqnarray}
f(x, \theta, \bar{\theta})=f(y)-i\theta\sigma^m\bar{\theta}\partial_m f(y)+\frac{1}{4}\theta\theta\bar{\theta}\bar{\theta}\Box f(y),
\end{eqnarray}
we always rewrite the results in the $(y, \theta, \bar{\theta})$ coordinate or $(x, \theta, \bar{\theta})$ either. For example,
\begin{eqnarray}
w_{\alpha}(x, \theta, \bar{\theta})&=&-4i\lambda_{\alpha}+4\theta_{\alpha}d-2i(\sigma^m \bar{\sigma}^k \theta)_{\alpha}v_{mk}\nonumber\\
&&+4\theta\theta\{\sigma^m\mathscr{D}_m\bar{\lambda}\}_\alpha+4(\theta\sigma^m\bar{\theta})\partial_m \lambda_\alpha\nonumber\\
&&
+2\theta\theta(\sigma^m\bar{\theta})_\alpha \{-i\partial_m d+\partial_m\partial^k v_k-\Box v_m\}   \nonumber\\
&& +\frac{i}{2}\theta\theta(\sigma^m\bar{\sigma}^k\sigma^l\bar{\theta})_\alpha\partial_l[v_k, v_m]
-i\theta\theta\bar{\theta}\bar{\theta}\Box\lambda_\alpha\label{W(x)}.
\end{eqnarray}
Note that they are not covariant under the super gauge transformation, because we take the WZ gauge fixing. Using \eref{DDD}, we obtain the result of calculation of $D^\alpha w_\alpha$ which is first term of the R.H.S of  \eref{Aterm} as
\begin{eqnarray}
D^\alpha w_\alpha(y, \theta, \bar{\theta})&=&-8d+8\theta\sigma^m\mathscr{D}_m\bar{\lambda}-8\bar{\theta}\bar{\sigma}^m\partial_m\lambda\nonumber\\
&&-8i(\bar{\theta}\bar{\sigma}^m \theta)\partial_m d-4(\bar{\theta}\bar{\sigma}^l\sigma^m\bar{\sigma}^k\theta)\partial_l v_{mk}\nonumber\\
&&-8i\theta\theta\{\bar{\theta}\bar{\sigma}^l\sigma^m\partial_l\mathscr{D}_m\bar{\lambda}\}.
\end{eqnarray}

%It is covariant under the residual super gauge transformation which means the transformation restricted under WZ gauge. 

%When we calculate $w_\alpha$ in terms of $(x, \theta, \bar{\theta})$ coordinates, the results is

%It seems not covariant under the super gauge transformation, because we take the WZ gauge fixing.  To clarify the covariance explicitly, it is useful to describe the result in terms of $(y, \theta, \bar{\theta})$ coordinates.
%Using the expansion relation  

%we obtain the results as 

%and it expresses the covariance under the residual super gauge transformation explicitly. Note that we can go back the coordinates $(x, \theta, \bar{\theta})$ every time using the relation 

\subsection{Useful Formulae}
We also give useful formulae to obtain the second term with component field of R.H.S. of \eref{Aterm} in terms of $(y, \theta, \bar{\theta})$ coordinates as
\begin{eqnarray}
e^{-\cv}D^{\alpha}e^{\cv}(y, \theta, \bar{\theta})&=&(\bar{\theta}\bar{\sigma}^m)^{\alpha}v_m+2i\theta^{\alpha}\bar{\theta}\bar{\lambda}-i\bar{\theta}\bar{\theta}\lambda^{\alpha}\nonumber\\
&&+\bar{\theta}\bar{\theta}(\theta^{\alpha}d-\frac{i}{2}(\theta\sigma^m\bar{\sigma}^k)^{\alpha}v_{km})\nonumber\\
&&-\theta\theta\bar{\theta}\bar{\theta}\mathscr{D}_m(\bar{\lambda}\bar{\sigma}^m)^{\alpha}.
\end{eqnarray}
%\begin{eqnarray}
%e^{-v}D_{\alpha}e^{v}(y, \theta, \bar{\theta})&=&-(\sigma^m\bar{\theta})_{\alpha} v_m+2i\theta_{\alpha}\bar{\theta}\bar{\lambda}-i\bar{\theta}\bar{\theta}\lambda_{\alpha}\nonumber\\
%&&+\bar{\theta}\bar{\theta}(\theta_{\alpha} d-\frac{i}{2}(\sigma^m\bar{\sigma}^k\theta)_{\alpha} v_{mk})\nonumber\\
%&&+\theta\theta\bar{\theta}\bar{\theta}(\sigma^m\mathscr{D}_m\bar{\lambda})_{\alpha}.
%\end{eqnarray}
Finally we obtain the $A$ in terms of $(y, \theta, \bar{\theta})$ coordinates as
%The right handed part of the main term in terms of $(y, \theta, \bar{\theta})$ coordinates is
\begin{eqnarray}
&&\hspace{-2cm}\left(D^\alpha w_\alpha+\{e^{-\cv}D^\alpha e^\cv, w_\alpha \}\right)(y, \theta, \bar{\theta})\nonumber\\
&=&-8d+8\theta\sigma^m\mathscr{D}_m\bar{\lambda}-8\bar{\theta}\bar{\sigma}^m\mathscr{D}_m\lambda\nonumber\\
&&+8[\bar{\theta}\bar{\lambda}, \theta\lambda]-8i(\bar{\theta}\bar{\sigma}^m \theta)\mathscr{D}_m d\nonumber\\
&&+4(\theta\sigma^k\bar{\sigma}^m\sigma^l\bar{\theta})\mathscr{D}_l v_{mk}\nonumber\\
&&-8i\theta\theta(\bar{\theta}\bar{\sigma}^l\sigma^m\mathscr{D}_l\mathscr{D}_m\bar{\lambda})\nonumber\\
&&+8i\theta\theta[\bar{\theta}\bar{\lambda}, d].
\end{eqnarray}
The $A^\dagger$ in terms of $(y^\dagger, \theta, \bar{\theta})$ coordinates is
\begin{eqnarray}
&&\hspace{-2cm}\left(D^\alpha w_\alpha+\{e^{-\cv}D^\alpha e^\cv, w_\alpha \}\right)^{\dagger}(y^\dagger, \theta, \bar{\theta})\nonumber\\
&=&-8d-8\bar{\theta}\bar{\sigma}^m\mathscr{D}_m{\lambda}+8{\theta}{\sigma}^m\mathscr{D}_m\bar{\lambda}\nonumber\\
&&+8[\bar{\theta}\bar{\lambda}, \theta\lambda]+8i(\bar{\theta}\bar{\sigma}^m \theta)\mathscr{D}_m d\nonumber\\
&&+4(\theta\sigma^l\bar{\sigma}^m\sigma^k\bar{\theta})\mathscr{D}_l v_{mk}\nonumber\\
&&+8i\bar{\theta}\bar{\theta}(\mathscr{D}_l\mathscr{D}_m{\lambda}\sigma^m\bar{\sigma}^l\theta)\nonumber\\
&&+8i\bar{\theta}\bar{\theta}[\lambda(y^{\dagger})\theta, d].
\end{eqnarray}

%%%%%%%%%%%%%%%%%%%%%%%%%%%%%%%%%%%%%%%%%%%%%%%%%%%%%%%%%%%%%%%%%%%%%%%%%%%%%%%%%%

\providecommand{\href}[2]{#2}\begingroup\raggedright\endgroup


\begin{thebibliography}{10}

\bibitem{Luscher:2010iy}
M.~Luscher, ``{Properties and uses of the Wilson flow in lattice QCD},''
  \href{http://dx.doi.org/10.1007/JHEP08(2010)071}{{\em JHEP} {\bfseries 1008}
  (2010) 071},
\href{http://arxiv.org/abs/1006.4518}{{\ttfamily arXiv:1006.4518 [hep-lat]}}.
%%CITATION = ARXIV:1006.4518;%%.

\bibitem{Luscher:2011bx}
M.~Luscher and P.~Weisz, ``{Perturbative analysis of the gradient flow in
  non-abelian gauge theories},''
  \href{http://dx.doi.org/10.1007/JHEP02(2011)051}{{\em JHEP} {\bfseries 1102}
  (2011) 051},
\href{http://arxiv.org/abs/1101.0963}{{\ttfamily arXiv:1101.0963 [hep-th]}}.
%%CITATION = ARXIV:1101.0963;%%.

\bibitem{Luscher:2013vga}
M.~Luscher, ``{Future applications of the Yang-Mills gradient flow in lattice
  QCD},''
\href{http://arxiv.org/abs/1308.5598}{{\ttfamily arXiv:1308.5598 [hep-lat]}}.
%%CITATION = ARXIV:1308.5598;%%.

\bibitem{Luscher:2013cpa}
M.~Luscher, ``{Chiral symmetry and the Yang--Mills gradient flow},''
  \href{http://dx.doi.org/10.1007/JHEP04(2013)123}{{\em JHEP} {\bfseries 1304}
  (2013) 123},
\href{http://arxiv.org/abs/1302.5246}{{\ttfamily arXiv:1302.5246 [hep-lat]}}.
%%CITATION = ARXIV:1302.5246;%%.

\bibitem{Suzuki:2013gza}
H.~Suzuki, ``{Energy-momentum tensor from the Yang-Mills gradient flow},''
  \href{http://dx.doi.org/10.1093/ptep/ptt059}{{\em PTEP} {\bfseries 2013}
  no.~8, (2013) 083B03},
\href{http://arxiv.org/abs/1304.0533}{{\ttfamily arXiv:1304.0533 [hep-lat]}}.
%%CITATION = ARXIV:1304.0533;%%.

\bibitem{DelDebbio:2013zaa}
L.~Del~Debbio, A.~Patella, and A.~Rago, ``{Space-time symmetries and the
  Yang-Mills gradient flow},''
  \href{http://dx.doi.org/10.1007/JHEP11(2013)212}{{\em JHEP} {\bfseries 1311}
  (2013) 212},
\href{http://arxiv.org/abs/1306.1173}{{\ttfamily arXiv:1306.1173 [hep-th]}}.
%%CITATION = ARXIV:1306.1173;%%.

\bibitem{Fodor:2012td}
Z.~Fodor, K.~Holland, J.~Kuti, D.~Nogradi, and C.~H. Wong, ``{The Yang-Mills
  gradient flow in finite volume},''
  \href{http://dx.doi.org/10.1007/JHEP11(2012)007}{{\em JHEP} {\bfseries 1211}
  (2012) 007},
\href{http://arxiv.org/abs/1208.1051}{{\ttfamily arXiv:1208.1051 [hep-lat]}}.
%%CITATION = ARXIV:1208.1051;%%.

\bibitem{Fodor:2012qh}
Z.~Fodor, K.~Holland, J.~Kuti, D.~Nogradi, and C.~H. Wong, ``{The gradient flow
  running coupling scheme},'' {\em PoS} {\bfseries LATTICE2012} (2012) 050,
\href{http://arxiv.org/abs/1211.3247}{{\ttfamily arXiv:1211.3247 [hep-lat]}}.
%%CITATION = ARXIV:1211.3247;%%.

\bibitem{Fritzsch:2013je}
P.~Fritzsch and A.~Ramos, ``{The gradient flow coupling in the Schr\"odinger
  Functional},'' \href{http://dx.doi.org/10.1007/JHEP10(2013)008}{{\em JHEP}
  {\bfseries 1310} (2013) 008},
\href{http://arxiv.org/abs/1301.4388}{{\ttfamily arXiv:1301.4388 [hep-lat]}}.
%%CITATION = ARXIV:1301.4388;%%.

\bibitem{Fritzsch:2013hda}
P.~Fritzsch and A.~Ramos, ``{Studying the gradient flow coupling in the
  Schr\"odinger functional},''
\href{http://arxiv.org/abs/1308.4559}{{\ttfamily arXiv:1308.4559 [hep-lat]}}.
%%CITATION = ARXIV:1308.4559;%%.

\bibitem{Ramos:2013gda}
A.~Ramos, ``{The gradient flow in a twisted box},''
\href{http://arxiv.org/abs/1308.4558}{{\ttfamily arXiv:1308.4558 [hep-lat]}}.
%%CITATION = ARXIV:1308.4558;%%.

\bibitem{Asakawa:2013laa}
{\bfseries FlowQCD Collaboration} Collaboration, M.~Asakawa, T.~Hatsuda,
  E.~Itou, M.~Kitazawa, and H.~Suzuki, ``{Thermodynamics of SU(3) Gauge Theory
  from Gradient Flow},''
\href{http://arxiv.org/abs/1312.7492}{{\ttfamily arXiv:1312.7492 [hep-lat]}}.
%%CITATION = ARXIV:1312.7492;%%.

\bibitem{Luscher:2014kea}
M.~Luscher, ``{Step scaling and the Yang-Mills gradient flow},''
  \href{http://dx.doi.org/10.1007/JHEP06(2014)105}{{\em JHEP} {\bfseries 1406}
  (2014) 105},
\href{http://arxiv.org/abs/1404.5930}{{\ttfamily arXiv:1404.5930 [hep-lat]}}.
%%CITATION = ARXIV:1404.5930;%%.

\bibitem{Rantaharju:2013bva}
J.~Rantaharju, ``{The Gradient Flow Coupling in Minimal Walking Technicolor},''
  {\em PoS} {\bfseries Lattice2013} (2014) 084,
\href{http://arxiv.org/abs/1311.3719}{{\ttfamily arXiv:1311.3719 [hep-lat]}}.
%%CITATION = ARXIV:1311.3719;%%.

\bibitem{Makino:2014wca}
H.~Makino and H.~Suzuki, ``{Lattice energy-momentum tensor from the Yang-Mills
  gradient flow -- a simpler prescription},''
\href{http://arxiv.org/abs/1404.2758}{{\ttfamily arXiv:1404.2758 [hep-lat]}}.
%%CITATION = ARXIV:1404.2758;%%.
%\if0{
\bibitem{Makino:2014taa}
H.~Makino and H.~Suzuki, ``{Lattice energy-momentum tensor from the Yang-Mills
  gradient flow -- inclusion of fermion fields},''
  \href{http://dx.doi.org/10.1093/ptep/ptu070}{{\em PTEP} 
  {\bfseries 2014} no.~6, (2014) 063B02},
\href{http://arxiv.org/abs/1403.4772}{{\ttfamily arXiv:1403.4772 [hep-lat]}}.
%%CITATION = ARXIV:1403.4772;%%.
%}\fi
\bibitem{Fritzsch:2013yxa}
P.~Fritzsch, A.~Ramos, and F.~Stollenwerk, ``{Critical slowing down and the
  gradient flow coupling in the Schr\"odinger functional},'' {\em PoS}
  {\bfseries Lattice2013} (2013) 461,
\href{http://arxiv.org/abs/1311.7304}{{\ttfamily arXiv:1311.7304 [hep-lat]}}.
%%CITATION = ARXIV:1311.7304;%%.

\bibitem{Bar:2013ora}
O.~Bar and M.~Golterman, ``{Chiral perturbation theory for gradient flow
  observables},'' \href{http://dx.doi.org/10.1103/PhysRevD.89.034505}{{\em
  Phys.Rev.} {\bfseries D89} (2014) 034505},
\href{http://arxiv.org/abs/1312.4999}{{\ttfamily arXiv:1312.4999 [hep-lat]}}.
%%CITATION = ARXIV:1312.4999;%%.

\bibitem{Brida:2013mva}
M.~Dalla~Brida and D.~Hesse, ``{Numerical Stochastic Perturbation Theory and
  the Gradient Flow},'' {\em PoS} {\bfseries Lattice2013} (2013) 326,
\href{http://arxiv.org/abs/1311.3936}{{\ttfamily arXiv:1311.3936 [hep-lat]}}.
%%CITATION = ARXIV:1311.3936;%%.

\bibitem{Monahan:2013lwa}
C.~Monahan and K.~Orginos, ``{Finite volume renormalization scheme for
  fermionic operators},'' {\em PoS} {\bfseries Lattice2013} (2013) 443,
\href{http://arxiv.org/abs/1311.2310}{{\ttfamily arXiv:1311.2310 [hep-lat]}}.
%%CITATION = ARXIV:1311.2310;%%.

\bibitem{Shindler:2013bia}
A.~Shindler, ``{Chiral Ward identities, automatic O(a) improvement and the
  gradient flow},''
  \href{http://dx.doi.org/10.1016/j.nuclphysb.2014.01.022}{{\em Nucl.Phys.}
  {\bfseries B881} (2014) 71--90},
\href{http://arxiv.org/abs/1312.4908}{{\ttfamily arXiv:1312.4908 [hep-lat]}}.
%%CITATION = ARXIV:1312.4908;%%.

\bibitem{wess}
J.~Wess and J.~Bagger, ``{Supersymmetry and Supergravity SECOND EDITION,
  REVISED AND EXPANDED},''
{\em Princeton Serieis in Physics} Princeton University Press, Princeton, New
  Jersey (1992).
%%CITATION = ARXIV:1302.5246;%%.



\bibitem{graham1}
G.~Robert, ``{Covariant Formulation of Non-Equilibrium Statistical
  Thermodynamics},'' {\em Z. Physik} {\bfseries B26} (1977) 397.

\bibitem{graham2}
G.~Robert, ``{COVARIANT STOCHASTIC CALCULUS IN THE SENSE OF IT$\mathrm{\hat{O}}$},''
  {\em PHYSICS LETTERS} {\bfseries 109A} (1985) 209.

\bibitem{Namiki:1984xe}
M.~Namiki, I.~Ohba, and K.~Okano, ``{Stochastic Quantization of Constrained
  Systems: General Theory and Nonlinear $\sigma$ Model},''
\href{http://dx.doi.org/10.1143/PTP.72.350}{{\em Prog.Theor.Phys.} {\bfseries
  72} (1984) 350}.
%%CITATION = PTPKA,72,350;%%.

\bibitem{Rumpf:1985eh}
H.~Rumpf, ``{Stochastic Quantization of Einstein Gravity},''
\href{http://dx.doi.org/10.1103/PhysRevD.33.942}{{\em Phys.Rev.} {\bfseries
  D33} (1986) 942}.
%%CITATION = PHRVA,D33,942;%%.

\bibitem{ZinnJustin:1986eq}
J.~Zinn-Justin, ``{Renormalization and Stochastic Quantization},''
\href{http://dx.doi.org/10.1016/0550-3213(86)90592-4}{{\em Nucl.Phys.}
  {\bfseries B275} (1986) 135}.
%%CITATION = NUPHA,B275,135;%%.

\bibitem{Halpern:1987ub}
M.~Halpern, ``{COORDINATE INVARIANT REGULARIZATION},''
\href{http://dx.doi.org/10.1016/0003-4916(87)90017-0}{{\em Annals Phys.}
  {\bfseries 178} (1987) 272}.
%%CITATION = APNYA,178,272;%%.

\bibitem{Nakazawa:1988ss}
N.~Nakazawa and D.~Ennyu, ``{Background Field Method for Nonlinear $\sigma$
  Model in Stochastic Quantization},''
\href{http://dx.doi.org/10.1016/0550-3213(88)90079-X}{{\em Nucl.Phys.}
  {\bfseries B305} (1988) 516}.
%%CITATION = NUPHA,B305,516;%%.

\bibitem{Nakazawa:1989jf}
N.~Nakazawa, ``{The Extended Local Gauge Invariance and the {BRS} Symmetry in
  Stochastic Quantization of Gauge Fields},''
\href{http://dx.doi.org/10.1016/0550-3213(90)90517-H}{{\em Nucl.Phys.}
  {\bfseries B335} (1990) 546}.
%%CITATION = NUPHA,B335,546;%%.

\bibitem{Nakazawa:2004ac}
N.~Nakazawa, ``{N = 1 supersymmetric Yang-Mills theory in Ito calculus},''
\href{http://dx.doi.org/10.1143/PTP.110.1117}{{\em Prog.Theor.Phys.} {\bfseries
  110} (2004) 1117--1150}.
%%CITATION = PTPKA,110,1117;%%.

\bibitem{Nakazawa:2003tz}
N.~Nakazawa, ``{Stochastic gauge fixing in N=1 supersymmetric Yang-Mills
  theory},'' \href{http://dx.doi.org/10.1143/PTP.116.883}{{\em
  Prog.Theor.Phys.} {\bfseries 116} (2007) 883--917},
\href{http://arxiv.org/abs/hep-th/0308081}{{\ttfamily arXiv:hep-th/0308081
  [hep-th]}}.
%%CITATION = HEP-TH/0308081;%%.


\end{thebibliography}
\end{document}